\documentclass[10pt,twocolumn,showpacs,preprintnumbers,amsmath,amssymb,aps,prl,superscriptaddress,longbibliography]{revtex4-2}
\usepackage{appendix}
\usepackage{bbm}
\usepackage{mathrsfs}
\usepackage{graphicx}
\usepackage{dcolumn}
\usepackage{bm}
\usepackage{braket}
\usepackage{amsmath}
\usepackage{amsfonts}
\usepackage[dvipsnames]{xcolor}
\usepackage[colorlinks=true,linkcolor=Blue,urlcolor=BlueViolet,citecolor=BlueViolet]{hyperref}
\usepackage{natbib}
\usepackage{multirow}

\begin{document}

\title{Coherent Information Phase Transition in a Noisy Quantum Circuit}
\author{Dongheng Qian}
\affiliation{State Key Laboratory of Surface Physics and Department of Physics, Fudan University, Shanghai 200433, China}
\affiliation{Shanghai Research Center for Quantum Sciences, Shanghai 201315, China}
\author{Jing Wang}
\thanks{Contact author: wjingphys@fudan.edu.cn}
\affiliation{State Key Laboratory of Surface Physics and Department of Physics, Fudan University, Shanghai 200433, China}
\affiliation{Shanghai Research Center for Quantum Sciences, Shanghai 201315, China}
\affiliation{Institute for Nanoelectronic Devices and Quantum Computing, Fudan University, Shanghai 200433, China}
\affiliation{Hefei National Laboratory, Hefei 230088, China}

\begin{abstract}

Coherent information quantifies the transmittable quantum information through a channel and is directly linked to the channel's quantum capacity. In a monitored quantum circuit, regarded as a quantum channel, extensive and positive coherent information is sustained at low measurement rates, protected by the scrambling dynamics. However, noise suppresses coherent information, driving it to zero or negative values. Here, we show that incorporating quantum-enhanced operations facilitates reliable quantum information transmission even in the presence of noise, as evidenced by a phase transition in coherent information from a recoverable phase with positive values to an irrecoverable phase with negative values. We provide both analytical understanding and numerical evidence demonstrating this transition, which is modulated by the relative frequencies of noise and quantum-enhanced operations. Additionally, we propose a resource-efficient protocol to characterize this phase transition in experiments, effectively avoiding post-selection by utilizing every run of the quantum circuit. This approach bridges the gap between theoretical insights and practical implementation, making the phase transition feasible to demonstrate on realistic noisy intermediate-scale quantum devices.

\end{abstract}

\date{\today}

\maketitle

Quantum information is fundamentally represented by quantum entanglement, which serves as a critical resource in both quantum computation and quantum communication~\cite{horodecki2009}. However, the transmission of quantum information is inevitably disrupted by interactions with the surrounding environment, leading to decoherence, which diminishes the potential advantages of quantum systems~\cite{nielsen2010, Carlesso2024}. To quantify the extent of information loss during quantum channel transmission, coherent information is a key metric, closely related to the quantum channel capacity~\cite{schumacher1996a, schumacher2002, horodecki2006a, schumacher1996b, wilde2013}. Positive coherent information indicates the successful transmission of finite quantum information through a channel, whereas zero or negative values suggest that no quantum information is being transmitted. As such, the development of methods to maintain positive coherent information in the presence of noise is a pivotal area of research in the quest for fault-tolerant quantum computation~\cite{gottesman2009}.

One approach to achieving positive coherent information involves encoding quantum information within an enlarged Hilbert space, transmitting it through a noisy quantum channel, and subsequently decoding it$-$this encapsulates the essence of quantum error correction~\cite{dennis2002, wang2003, kitaev1997, terhal2015, kitaev2003a}.
An alternative method involves encoding information in a highly non-local manner within the same Hilbert space, leveraging quantum scrambling~\cite{sekino2008, mi2021, landsman2019}. In the context of measurement-induced phase transitions (MIPT), it has been shown that a low rate of local measurements is insufficient to extract significant information when competing with scrambling dynamics generated by random unitary gates~\cite{li2019, chan2019, bao2024, lee2022, li2023a, nahum2020, nahum2018, nahum2019, sharma2022, sang2021a, skinner2019, szyniszewski2019, vasseur2019, zabalo2020, alberton2021, fidkowski2021, fisher2023, poboiko2024, yu2022, choi2020}. This phenomenon is more transparently elucidated by relating MIPT to a dynamical purification transition~\cite{gullans2020a}, characterized by circuit-averaged coherent information, where the input state is a completely mixed state. In such scenarios, coherent information is extensive in system size and remains positive in the mixed phase, while it approaches zero in the pure phase. However, it is important to note that scrambling alone is insufficient to protect information transmission from other prevalent sources of noise and may even aggravate the suppression of entanglement~\cite{dias2023, liu2023e, liu2024, liu2024a, weinstein2022, Haas2024, Weinstein2024, lovas2024}. As a result, negative coherent information is expected in noisy quantum circuits involving random unitary gates and measurements.

In this Letter, we explore the integration of quantum-enhanced (QE) operations~\cite{braun2018, aharonov2022, huang2021b, zhao2017, kelly2024a, kelly2024b, qian2024b} into quantum circuits, unveiling a phase transition in coherent information that is governed by the relative frequency of various noises and QE operations. In the recoverable phase, the coherent information remains positive, indicating reliable quantum information transmission even in the presence of noise, while in the irrecoverable phase, it becomes negative. Our results thus demonstrate that QE operations are effective not only in storing quantum information~\cite{qian2024b}, but also in transmitting it. This transition can be analytically understood by mapping the circuit to a classical statistical mechanics model and is explicitly demonstrated through numerical simulations. Moreover, we demonstrate that this phase transition can be efficiently probed in experiments by generalizing the cross-entropy benchmark for MIPT~\cite{li2023}, thereby circumventing the challenges associated with post-selection. This finding underscores that the phase transition remains experimentally tractable even in the thermodynamic limit, where the phases of matter are well defined~\cite{friedman2023}.

\begin{figure}[t]
\begin{center}
\includegraphics[width=3.4in, clip=true]{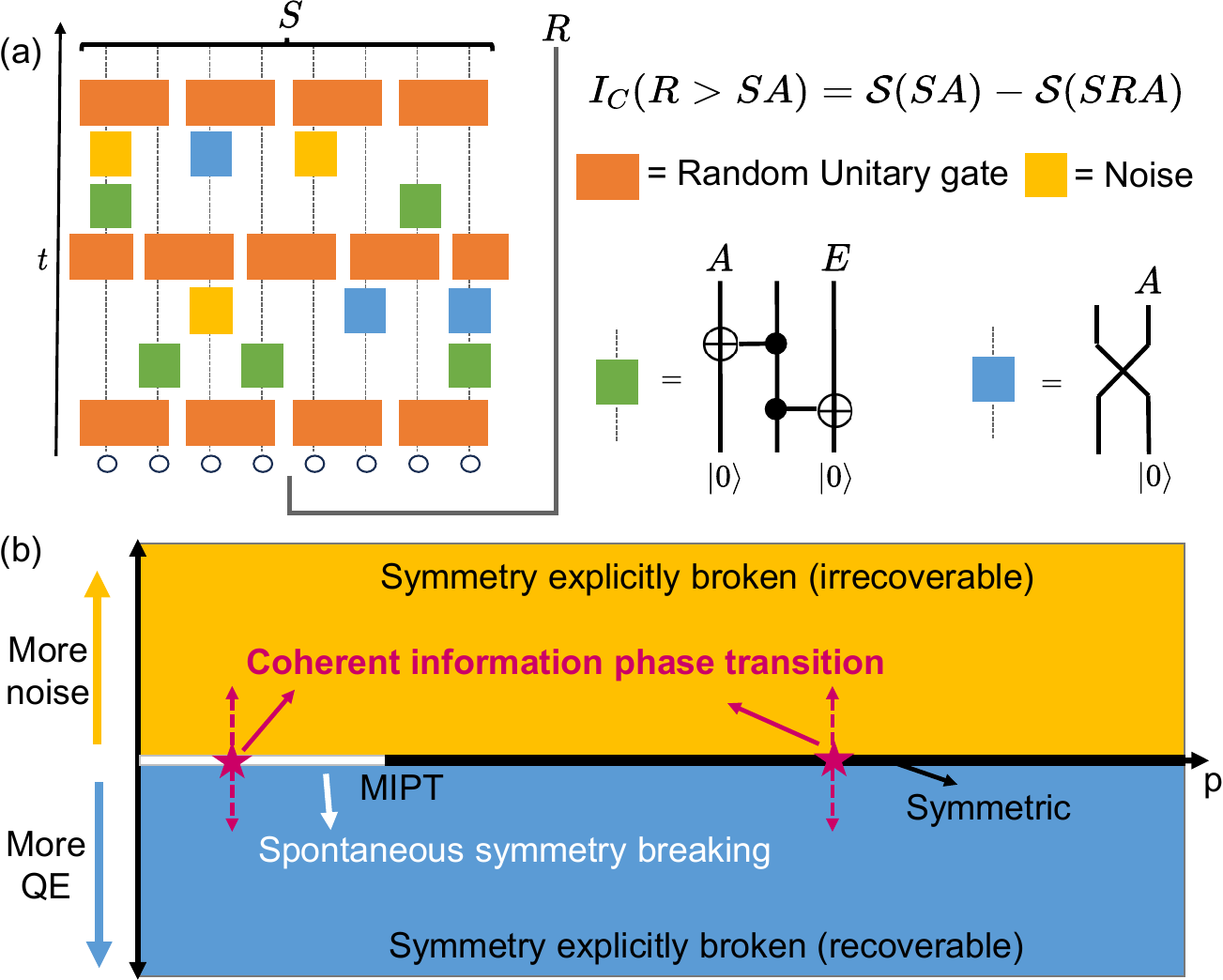}
\end{center}
\caption{Circuit structure and phase diagram. (a) Circuit structure. Orange, green, yellow, and blue rectangles represent unitary gates, measurements, noise, and QE operations, respectively. The initial state is entangled with reference qubits $R$. We primarily consider QE operations where $U_{SA} = \text{SWAP}$. (b) Phase diagram. We focus on the coherent information phase transition from recoverable to irrecoverable phase, tuned by the relative frequency of noise and QE operations. This phase transition manifests irrespective of the measurement rate $p$.}
\label{fig1}
\end{figure}

\emph{Circuit model---}We consider a quantum circuit structure composed of four types of operations, as depicted in Fig.~\ref{fig1}(a). Random unitary gates are applied in a brick-wall pattern. Between each pair of unitary layers, each qubit has a probability $p$ of being projectively measured along the $z$-axis. The projective measurement can be modeled by first applying a $\text{CNOT}$ gate between the system qubit and an environment qubit, followed by another $\text{CNOT}$ gate with an ancilla qubit. Tracing out the environment qubit leaves the classical measurement outcome in the ancilla qubit. After the measurement, each qubit has a probability $q_n$ of undergoing a certain noise channel, such as depolarizing, resetting, or dephasing, and a probability $q_e$ of undergoing a QE operation. We denote $q_t = q_n + q_e$ and $q = q_n / q_t$. QE operations involve utilizing a quantum probe to extract information from the system, which can be conceptualized as a dynamic expansion of the Hilbert space through the introduction of ancilla qubits during the circuit evolution. Specifically, a QE operation introduces ancilla qubits and applies a unitary gate to the system qubit and these ancilla qubits, which are then isolated and left untouched until the end of the circuit. Given that each ancilla qubit is used only once, we assume no noise affects them during the evolution, which is a reasonable assumption considering they can be well-isolated from other qubits, thereby maintaining a coherence time much longer than that of the system qubits. 
% In Refs.~\cite{kelly2024a, kelly2024b, qian2024b}, the case where noise and QE operations are symmetric was studied. Here, we consider the relative frequency of noise and QE operations, $q$, is no longer constrained to be $0.5$.
In the following, we denote the system qubits undergoing evolution as $S$, the environment qubits as $E$, and the ancilla qubits as $A$.

The input state is entangled with a reference system $R$, and the entire quantum circuit can be viewed as a quantum channel from $R$ to $SA$, where $SA \equiv S \cup A$. Without loss of generality, we consider the scenario where each system qubit is entangled in a Bell pair with a corresponding reference qubit. It is important to note that $A$ includes both ancilla qubits from QE operations that store quantum information and those from measurements that only contain classical information. Coherent information, a key quantity for assessing quantum channel capacity, is computed as:
\begin{equation}
I_C(R>SA) = \mathcal{S}(SA)-\mathcal{S}(SRA),
\end{equation}
where $SRA \equiv S \cup R \cup A$ and $\mathcal{S}(X)$ denoted the entanglement entropy of the subsystem $X$. Our primary focus is to investigate whether a phase transition in $I_C(R>SA)$ occurs as $q$ is varied.

\emph{Analytical analysis---}When random unitary gates are drawn from the Haar measure, the coherent information can be mapped onto the free energy difference of a classical statistical mechanics model under different boundary conditions using the replica trick~\cite{bao2020, jian2020a, zhou2019, li2021a}. Specifically, we have
\begin{eqnarray}
I_C(R>SA) &=& \lim_{n \to 1}(\mathcal{S}^{(n)}(SA) - \mathcal{S}^{(n)}(SRA)) \nonumber\\
&=&\lim_{n \to 1}\lim_{k \to 0}\frac{1}{(1-n)k}\text{log}\left(\frac{\mathcal{Z}^{(n,k)}_{SA}}{\mathcal{Z}^{(n,k)}_{SRA}}\right)\\
&=&\lim_{n \to 1}\lim_{k \to 0}\frac{1}{(n-1)k}\left(\mathcal{F}^{(n,k)}_{SA}-\mathcal{F}^{(n,k)}_{SRA}\right),\nonumber
\label{eq:one}
\end{eqnarray}
where
\begin{eqnarray}
\mathcal{Z}_{XA}^{(n,k)} &=& \sum_{m}\text{Tr}\left(\mathbb{C}^{X}\Lambda^{(n,k)}\right), \mathcal{F}_{XA}^{(n,k)}=-\text{log}\mathcal{Z}_{XA}^{(n,k)}.
\label{eq:two}
\end{eqnarray}
$X=S, SR$. $\Lambda^{(n,k)}$ can be considered the bulk partition function, where the underlying degrees of freedom are spins that take values in the permutation group $\mathbb{S}(Q)$ with $Q=nk+1$. $\mathcal{Z}_{SA}^{(n,k)}$ and $\mathcal{Z}_{SRA}^{(n,k)}$ are interpreted as total partition functions under different boundary conditions. We denote $\mathbb{I}$ and $\mathbb{C}$ as the identity element and another specific group element in $\mathbb{S}(Q)$, constructed from swap operations, with its precise definition provided in the Supplemental Material~\cite{supp}. Specifically, $\mathbb{C}^{S}$ indicates that the top boundary aligns with $\mathbb{C}$ and the bottom with $\mathbb{I}$, while $\mathbb{C}^{SR}$ means both top and bottom boundaries align with $\mathbb{C}$.

The statistical mechanics model encoded in $\Lambda^{(n,k)}$ is a ferromagnetically coupled spin model with random fields oriented in two different directions on a honeycomb lattice~\cite{qian2024b}. The vertical bonds are determined by the random unitary gates, which act as a ferromagnetic coupling that assigns higher weights to configurations where neighboring spins are aligned in the same direction.
The non-vertical bonds are determined by measurements, QE operations and noise. Although their exact contributions depend on the specific QE operations and noise model, noises generally behave as external fields applied at random locations, aligning in the direction $\mathbb{I}$.
% , corresponding to the identity element in $\mathbb{S}(Q)$
In contrast, QE operation acts as an external field oriented in the direction $\mathbb{C}$. The parameter $q$ governs the relative frequency of QE operations and noise, thus tuning the direction of net external field. Concrete expressions for the bond weights under various noises are provided in~\cite{supp}. 

For small $q$, bulk spins align with $\mathbb{C}$, leading to a domain wall at the bottom boundary when calculating $\mathcal{F}^{(n,k)}_{SA}$, with no additional energy cost in $\mathcal{F}^{(n,k)}_{SRA}$. This results in a positive $I_C$ and is extensive in system size. Since positive $I_C$ indicates that finite amount of quantum information can be successfully transmitted through the channel, which can be recovered via a decoding algorithm, we term this the recoverable phase. Conversely, for large $q$, bulk spins align with $\mathbb{I}$. In this scenario, a domain wall at the top boundary always exists, while an extra domain wall appears in calculating $\mathcal{F}^{(n,k)}_{SRA}$ at the bottom boundary, rendering $I_C$ negative and resulting in an irrecoverable phase. 
This coherent information phase transition, tuned by $q$, exhibits characteristics of a first-order phase transition when the sign of $I_C$ is considered as the order parameter. This contrasts with the MIPT protected by QE operations, which corresponds to a second-order phase transition between ferromagnetic and paramagnetic phases~\cite{qian2024b}. The full phase diagram is depicted in Fig.~\ref{fig1}(b). By identifying $p$ as the temperature and considering noise and QE operations as two competing external fields, the phase diagram is reminiscent of that of a 2D Ising model. A central contribution of our work is the recognition that the additional symmetry-breaking phase induced by QE operations in the classical statistical mechanics mapping corresponds to the recoverable phase, in which quantum information can be reliably transmitted.

\begin{figure}[t]
\begin{center}
\includegraphics[width=3.4in, clip=true]{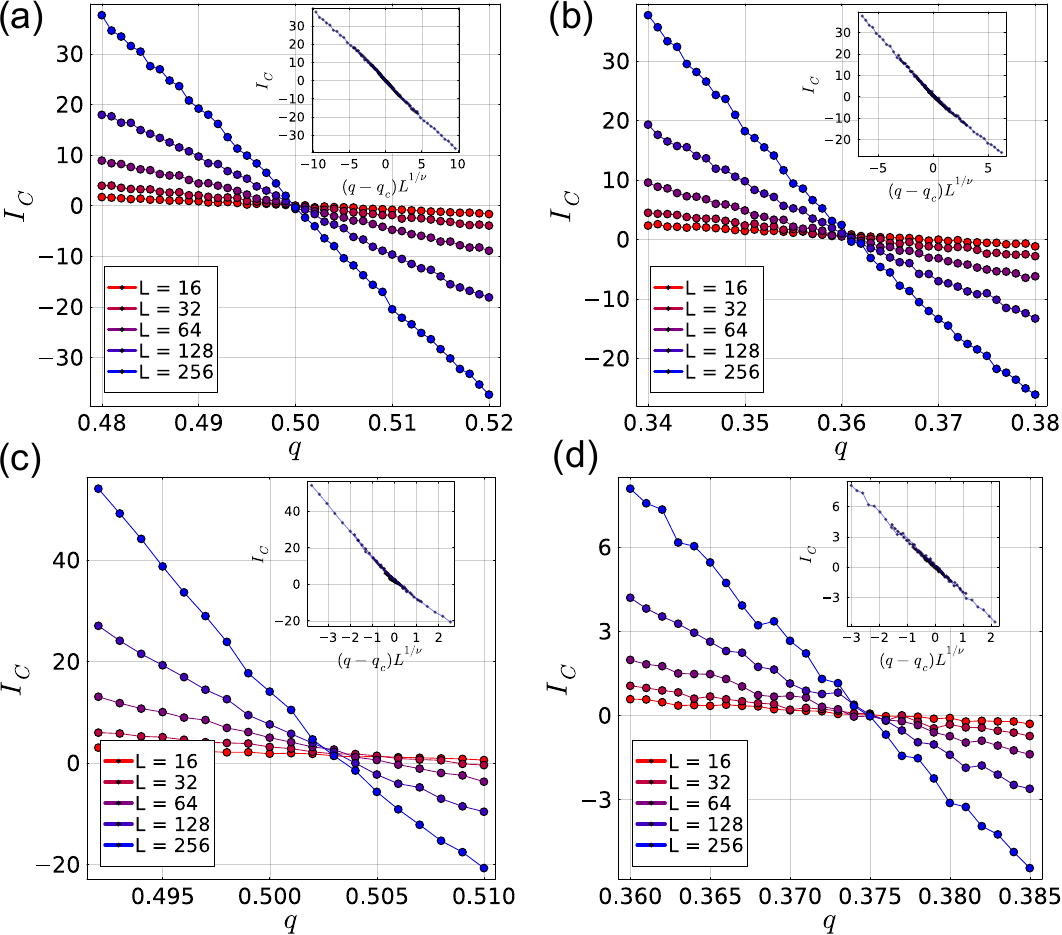}
\end{center}
\caption{Numerical simulation for coherent information phase transition. $L$ denotes number of qubits. (a) Resetting noise. (b) Depolarizing noise. (c) Dephasing noise. (d) Depolarizing noise with measurement probablility $p=0.1$. Insets show the data collapse results. Every data point is averaged over $6\times10^3$ realizations. The circuit is evolved for $5L$ time steps. }
\label{fig2}
\end{figure}

\emph{Numerical results---}To explicitly demonstrate the existence of the coherent information phase transition, we perform large-scale numerical simulations in a (1+1)-d quantum circuit. In these simulations, random unitary gates are selected from the Clifford group to enable the use of the stabilizer formalism~\cite{aaronson2004, gottesman1998, gottesman1997}.  We set $U_{SA} = \text{SWAP}$ and specifically consider three types of noise: resetting, depolarizing, and dephasing. Here we set the total frequency of noise and QE operations to be $q_t=0.1$, while results for other parameters can be found in~\cite{supp}.
We first consider the case with measurement probability $p=0$, and the results are presented in Fig.~\ref{fig2}(a-c). A distinct recoverable phase, characterized by positive coherent information, is observed in all cases, highlighting the effectiveness of QE operations in combating various noises. Through data collapse, we determine the critical points and critical exponents to be $q_c^{\text{reset}}=0.500(1)$, $q_c^{\text{depo}}=0.360(1)$ and $q_c^{\text{deph}}=0.503(1)$, with $\nu_c^{\text{reset}} = 0.90(5)$, $\nu_c^{\text{depo}}=0.96(5)$ and $\nu_c^{\text{deph}}=0.95(7)$, respectively.
The recoverable phase region is the smallest for depolarizing noise and the largest for dephasing noise, consistent with the relative strengths of the corresponding random fields~\cite{supp}. When measurements are incorporated, the phase transition persists, as demonstrated in Fig.~\ref{fig2}(d) for depolarizing noise, albeit at a different critical point $q_c = 0.375(1)$ with $\nu_c=1.03(8)$. Additional numerical details and results are provided in~\cite{supp}. 

\begin{figure}[t]
\begin{center}
\includegraphics[width=3.4in, clip=true]{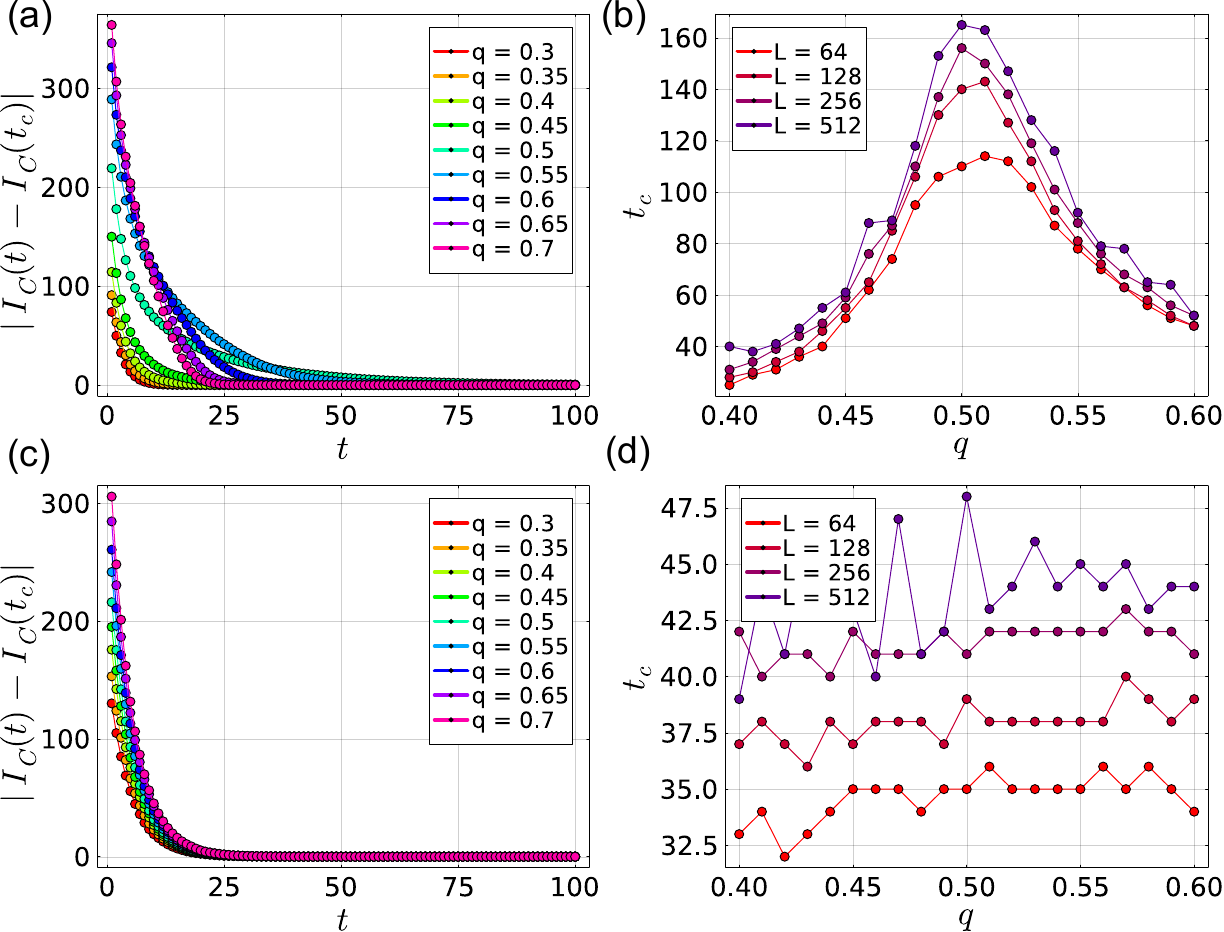}
\end{center}
\caption{Numerical evidence of critical slowing down. We consider resetting noise with $p=0$ as an example. (a) Temporal evolution of $I_C$ for varying $q$. We take system size to be $L=256$. (b) Convergence time vs. $q$ for multiple system sizes. (c), (d) analogous to (a), (b), but in absence of random unitary gates.}
\label{fig3}
\end{figure}

Another interesting phenomenon we observed is critical slowing down~\cite{marconi2024, scheffer2009, vanhove1954, li2022}. As illustrated in Fig.~\ref{fig3}(a), the convergence of coherent information slows significantly near $q=0.5$ for resetting noise. 
By defining the convergence time $t_c$ as the time when $\left | I_C(t)-I_C(t_\text{final}) \right |$ first falls below $0.05$, we plot $t_c$ as a function of $q$ for different system sizes in Fig.~\ref{fig3}(b). 
The results indicate that $t_c$ diverges with system size as $q$ approaches the critical point, suggesting an infinite convergence time in the thermodynamic limit. In contrast, in the absence of unitary gates, while coherent information still transitions from positive to negative precisely at $q_c = 0.5$ (corresponding to the scenario where half of the system qubits are discarded), critical slowing down was not observed, as shown in Fig.~\ref{fig3}(c) and Fig.~\ref{fig3}(d). This observation underscores the indispensability of unitary gates for the manifestation of a phase transition.

\emph{Efficient protocol---}Measuring coherent information for a specific circuit and trajectory requires repeated preparation of the same final state. However, in realistic experiments, the noise locations are usually uncontrollable, making it difficult to reliably reproduce the same circuit. This issue is compounded by the post-selection problem when measurements are involved, due to Born's rule~\cite{qian2024a, Paviglianiti2024, friedman2023}. 
Motivated by previous work on constructing post-selection-free probes that reflect the correlation between a quantum device and classical simulations~\cite{garratt2024, li2023, mcginley2024, tikhanovskaya2024, kamakari2024}, we propose a resource-efficient protocol to observe the coherent information phase transition, utilizing every run of the quantum circuit. It is important to note that our protocol relies on the knowledge of noise locations. While controlling where such noise occurs is inherently challenging, identifying its location post-occurrence is considerably more tractable.
Consider running the same circuit realization $\mathcal{C}$ twice, each with distinct initial states $\rho$ and $\sigma$. The circuit realization $\mathcal{C}$ includes choices of random unitary gates and locations of noise, QE operations and measurements. The unnormalized final state on ancilla qubits $A$ is denoted as $ \widetilde{\rho_A^{m}}$ and $\widetilde{\sigma_A^{m}}$, where $\mathrm{Tr}(\widetilde{\rho_A^{m}}) = p_{\rho}^{m}$ and $\mathrm{Tr}(\widetilde{\sigma_A^{m}}) = p_{\sigma}^{m}$, with $p_{\rho}^{m}$ and $p_{\sigma}^{m}$ representing the probability of the particular trajectory $m$ for each respective run. For brevity, we omit the explicit dependence of $\widetilde{\rho_A^{m}}$ and $\widetilde{\sigma_A^{m}}$ on $\mathcal{C}$. We can then define the following quantity:
\begin{equation}
\chi = \mathbb{E}_{\mathcal{C}}\frac{\sum_{m}\text{Tr}\left(\widetilde{\rho_{A}^{m}} \widetilde{\sigma_{A}^{m}}\right)}{\sum_{m}\text{Tr}\left(\widetilde{\sigma_{A}^{m}}^{2}\right)},
\end{equation}
where $\mathbb{E}_{\mathcal{C}}$ represents averaging over different circuit realizations. It resembles the ``quantum'' cross entropy introduced in Ref.~\cite{li2023}, with the notable distinction that ancilla qubits  introduced dynamically during the evolution, rather than being present from the beginning. Furthermore, while $\chi$ does not reveal MIPT, it is a robust indicator of the coherent information transition. Intuitively, this probe reflects the distinguishability of different initial states by leveraging access to both the measurement outcomes and the final state of the ancilla qubits. In the recoverable phase, the initial information is preserved, and therefore we expect $\chi=0$, corresponding to perfect distinguishability. In contrast, in the irrecoverable phase, the initial information is erased, causing $\chi$ to approach 1. A rigorous analytical analysis demonstrating the sharp transition of $\chi$ at the phase transition is provided in~\cite{supp}.

\begin{figure}[t]
\begin{center}
\includegraphics[width=3.4in, clip=true]{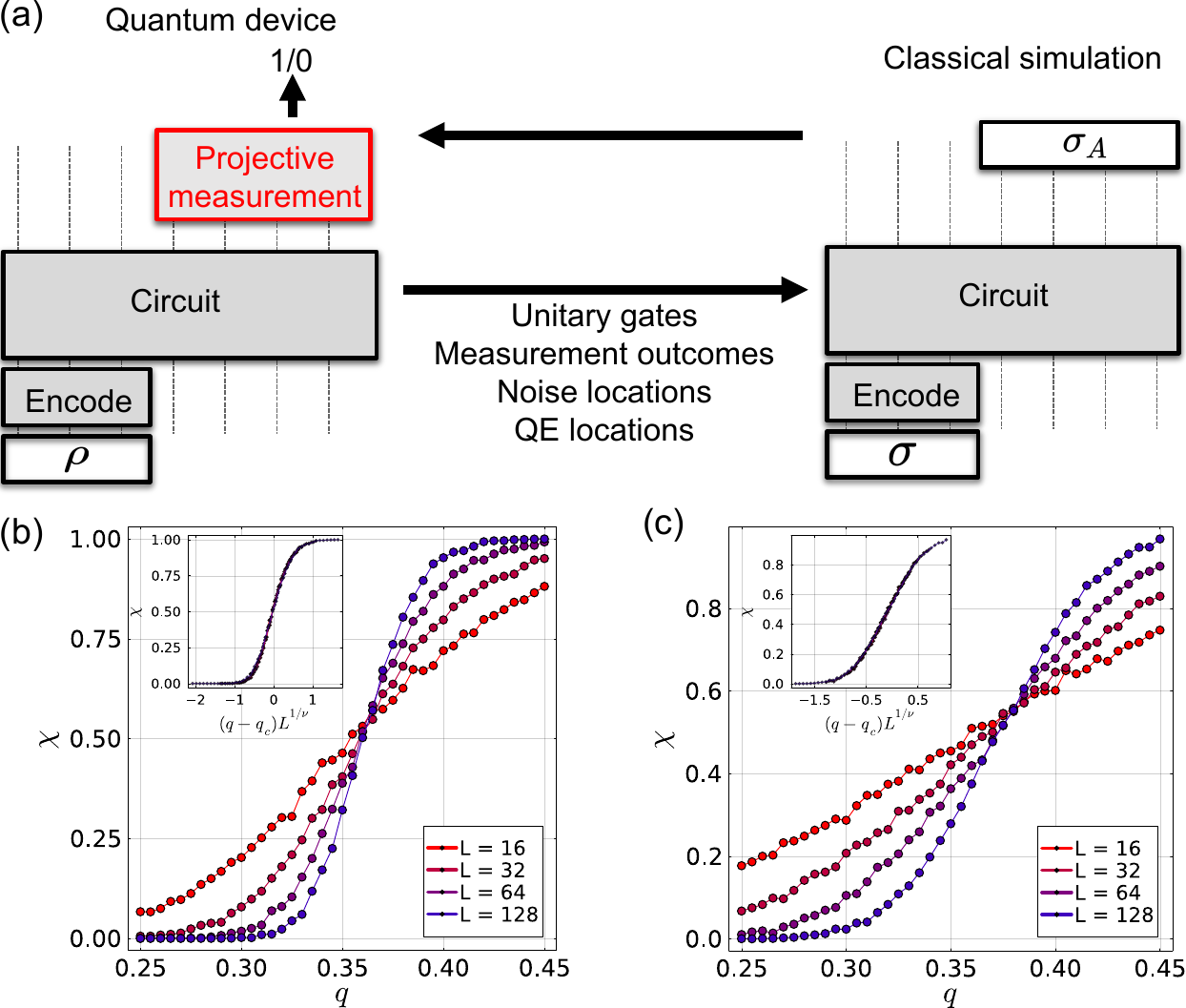}
\end{center}
\caption{Efficient protocol. (a) Schematic workflow of the protocol. An important additional ingredient, compared to the approach in Ref.~\cite{li2023}, is the incorporation of projective measurements on the ancilla qubits. (b) Take depolarizing noise as an example. We choose $ \rho = (\left | + \right \rangle \left \langle + \right \rangle)^{\otimes L/2}\otimes(\left | 0 \right \rangle \left \langle 0 \right \rangle)^{\otimes L/2}$ and $ \sigma = (\left | 0 \right \rangle \left \langle 0 \right \rangle)^{\otimes L}$. Every data point is averaged over $3\times10^3$ realizations. The inset shows the data collapse result. (c) Analogous to (b), with measurement probability $p=0.1$.}
\label{fig4}
\end{figure}

A key advantage of $\chi$ is that its measurement allows us to fully utilize each circuit realization and trajectory, circumventing the need for experimental control over noise locations or post-selection on measurement outcomes. The central idea is that the run with initial state $\sigma$ can be classically simulated by selecting it as a stabilizer state. When the unitary gates are Clifford gates, $\widetilde{\sigma_{A}^{m}}$ remains a stabilizer state, leading to the following expression for $\chi$:
\begin{equation}
\chi= \mathbb{E}_{\mathcal{C}}\sum_mp_{\rho}^{m}\frac{p_{\sigma}^{m}}{2^{-N_{\text{rand}}^{\sigma}}}\text{Tr}\left(\rho_A^{m}\prod_{k=1}^{l_{\sigma}}\frac{1+g_{k}}{2}\right).
\end{equation}
Here, $N_{\text{rand}}^{\sigma}$ is the number of measurements with random outcomes during the classical simulation, $l_{\sigma}$ represents the number of generators for the normalized stabilizer state $\sigma_{A}^{m}$ and $g$ are the corresponding generators. The middle term $p_{\sigma}^{m}/2^{-N_{\text{rand}}^{\sigma}}$ is either $0$ or $1$ due to the property of stabilizer formalism, while the final term corresponds to the probability of measuring all the generators with outcomes equal to $1$. Thus, the overall process for measuring $\chi$ is outlined in Fig.~\ref{fig4}(a) and is detailed as follows. For each execution of the quantum circuit on a quantum device, we first conduct the classical simulation of the same circuit with initial state $\sigma$, leveraging the knowledge of the measurement outcomes and the circuit realization. 
We then compute $p_{\sigma}^{m}/2^{-N_{\text{rand}}^{\sigma}}$ via classical simulation. If the result is $1$, we proceed to perform projective measurements on the ancilla qubits in the quantum device according to the stabilizer generators. If all measurement outcomes are $1$, we denote this as a successful event. The quantity $\chi$ is then estimated as the proportion of successful events among all circuit executions. 
To demonstrate the efficacy of this probe, we present numerical results in Fig.~\ref{fig4}(b) and Fig.~\ref{fig4}(c) for the case of depolarizing noise with $p = 0$ and $p = 0.1$, respectively. Here, $\rho$ is chosen as a stabilizer state for simulating the quantum device, but it could in principle be any state, rendering the evolution intractable classically~\cite{kamakari2024}. The critical points are determined to be $q_c = 0.362(2)$ and $q_c = 0.378(2)$, which are in agreement with the exact phase transition points. 

\emph{Discussions---}The coherent information is related to the single-shot quantum channel capacity by the relation $ \mathcal{Q}_1 = \max_{\rho_{SR}}I_C(R>SA)$~\cite{wilde2013, nielsen2010}. Here we consider $\rho_{SR}$ to be a direct product of Bell pairs between $S$ and $R$. Through analytical mapping, it is evident that this choice maximizes coherent information within the recoverable phase, minimizing the free energy $\mathcal{F}^{(n,k)}_{SAR}$. However, in the irrecoverable phase, selecting $\rho_{SR}$ as a trivial product state would yield zero coherent information, thereby maximizing the coherent information. This results in a phase transition where the single-shot quantum channel capacity shifts from a finite positive value to zero. 
Generalizing to the true quantum channel capacity $\mathcal{Q}$, which allows for multiple uses of the channel, general noises are non-degradable, making $\mathcal{Q}$ non-additive and complicating the calculation~\cite{holevo1998, devetak2005, hastings2009, devetak2005a, lloyd1997, divincenzo1998, barnum2000a}. Thus, it is intriguing to explore whether a phase transition in $\mathcal{Q}$ can be identified within this framework.

A key aspect of the resource-efficient protocol for detecting the coherent information phase transition is the requirement for knowledge of noise locations.
In a practical experimental setting, it has been demonstrated that various types of noise can be effectively converted into erasure errors, which can be located by verifying whether the qubit remains in the computational subspace~\cite{delfosse2020, solanki2023, grassl1997, bennett1997, chang2024, kang2023, wu2022a, kubica2023, gu2024}. Upon detection of an erasure error, replacing the qubit with a maximally mixed state corresponds to a depolarizing noise, while replacing it with $\left | 0 \right >$ corresponds to a resetting noise. It is therefore of interest to investigate whether imperfect detection of erasure errors would compromise the protocol's success. We leave this to future work.

\begin{acknowledgments}
\emph{Acknowledgment.} We thank Xiao-Liang Qi and Chong Wang for valuable discussions. This work is supported by the Natural Science Foundation of China through Grants No.~12350404 and No.~12174066, the Innovation Program for Quantum Science and Technology through Grant No.~2021ZD0302600, the Science and Technology Commission of Shanghai Municipality under Grants No.~23JC1400600, No.~24LZ1400100 and No.~2019SHZDZX01, and is sponsored by the ``Shuguang Program'' supported by the Shanghai Education Development Foundation and Shanghai Municipal Education Commission.
\end{acknowledgments}


\begin{thebibliography}{90}%
\makeatletter
\providecommand \@ifxundefined [1]{%
 \@ifx{#1\undefined}
}%
\providecommand \@ifnum [1]{%
 \ifnum #1\expandafter \@firstoftwo
 \else \expandafter \@secondoftwo
 \fi
}%
\providecommand \@ifx [1]{%
 \ifx #1\expandafter \@firstoftwo
 \else \expandafter \@secondoftwo
 \fi
}%
\providecommand \natexlab [1]{#1}%
\providecommand \enquote  [1]{``#1''}%
\providecommand \bibnamefont  [1]{#1}%
\providecommand \bibfnamefont [1]{#1}%
\providecommand \citenamefont [1]{#1}%
\providecommand \href@noop [0]{\@secondoftwo}%
\providecommand \href [0]{\begingroup \@sanitize@url \@href}%
\providecommand \@href[1]{\@@startlink{#1}\@@href}%
\providecommand \@@href[1]{\endgroup#1\@@endlink}%
\providecommand \@sanitize@url [0]{\catcode `\\12\catcode `\$12\catcode `\&12\catcode `\#12\catcode `\^12\catcode `\_12\catcode `\%12\relax}%
\providecommand \@@startlink[1]{}%
\providecommand \@@endlink[0]{}%
\providecommand \url  [0]{\begingroup\@sanitize@url \@url }%
\providecommand \@url [1]{\endgroup\@href {#1}{\urlprefix }}%
\providecommand \urlprefix  [0]{URL }%
\providecommand \Eprint [0]{\href }%
\providecommand \doibase [0]{https://doi.org/}%
\providecommand \selectlanguage [0]{\@gobble}%
\providecommand \bibinfo  [0]{\@secondoftwo}%
\providecommand \bibfield  [0]{\@secondoftwo}%
\providecommand \translation [1]{[#1]}%
\providecommand \BibitemOpen [0]{}%
\providecommand \bibitemStop [0]{}%
\providecommand \bibitemNoStop [0]{.\EOS\space}%
\providecommand \EOS [0]{\spacefactor3000\relax}%
\providecommand \BibitemShut  [1]{\csname bibitem#1\endcsname}%
\let\auto@bib@innerbib\@empty
%</preamble>
\bibitem [{\citenamefont {Horodecki}\ \emph {et~al.}(2009)\citenamefont {Horodecki}, \citenamefont {Horodecki}, \citenamefont {Horodecki},\ and\ \citenamefont {Horodecki}}]{horodecki2009}%
  \BibitemOpen
  \bibfield  {author} {\bibinfo {author} {\bibfnamefont {R.}~\bibnamefont {Horodecki}}, \bibinfo {author} {\bibfnamefont {P.}~\bibnamefont {Horodecki}}, \bibinfo {author} {\bibfnamefont {M.}~\bibnamefont {Horodecki}},\ and\ \bibinfo {author} {\bibfnamefont {K.}~\bibnamefont {Horodecki}},\ }\bibfield  {title} {\bibinfo {title} {Quantum entanglement},\ }\href {https://doi.org/10.1103/RevModPhys.81.865} {\bibfield  {journal} {\bibinfo  {journal} {Rev. Mod. Phys.}\ }\textbf {\bibinfo {volume} {81}},\ \bibinfo {pages} {865} (\bibinfo {year} {2009})}\BibitemShut {NoStop}%
\bibitem [{\citenamefont {Nielsen}\ and\ \citenamefont {Chuang}(2010)}]{nielsen2010}%
  \BibitemOpen
  \bibfield  {author} {\bibinfo {author} {\bibfnamefont {M.~A.}\ \bibnamefont {Nielsen}}\ and\ \bibinfo {author} {\bibfnamefont {I.~L.}\ \bibnamefont {Chuang}},\ }\href {https://doi.org/10.1017/CBO9780511976667} {\emph {\bibinfo {title} {Quantum computation and quantum information}}}\ (\bibinfo  {publisher} {Cambridge university press},\ \bibinfo {year} {2010})\BibitemShut {NoStop}%
\bibitem [{\citenamefont {Carlesso}(2024)}]{Carlesso2024}%
  \BibitemOpen
  \bibfield  {author} {\bibinfo {author} {\bibfnamefont {M.}~\bibnamefont {Carlesso}},\ }\href {https://arxiv.org/abs/2406.11613} {\bibinfo {title} {Lecture notes on quantum algorithms in open quantum systems}} (\bibinfo {year} {2024}),\ \Eprint {https://arxiv.org/abs/2406.11613} {arXiv:2406.11613 [quant-ph]} \BibitemShut {NoStop}%
\bibitem [{\citenamefont {Schumacher}\ and\ \citenamefont {Nielsen}(1996)}]{schumacher1996a}%
  \BibitemOpen
  \bibfield  {author} {\bibinfo {author} {\bibfnamefont {B.}~\bibnamefont {Schumacher}}\ and\ \bibinfo {author} {\bibfnamefont {M.~A.}\ \bibnamefont {Nielsen}},\ }\bibfield  {title} {\bibinfo {title} {Quantum data processing and error correction},\ }\href {https://doi.org/10.1103/PhysRevA.54.2629} {\bibfield  {journal} {\bibinfo  {journal} {Phys. Rev. A}\ }\textbf {\bibinfo {volume} {54}},\ \bibinfo {pages} {2629} (\bibinfo {year} {1996})}\BibitemShut {NoStop}%
\bibitem [{\citenamefont {Schumacher}\ and\ \citenamefont {Westmoreland}(2002)}]{schumacher2002}%
  \BibitemOpen
  \bibfield  {author} {\bibinfo {author} {\bibfnamefont {B.}~\bibnamefont {Schumacher}}\ and\ \bibinfo {author} {\bibfnamefont {M.~D.}\ \bibnamefont {Westmoreland}},\ }\bibfield  {title} {\bibinfo {title} {Approximate quantum error correction},\ }\href {https://link.springer.com/article/10.1023/A:1019653202562} {\bibfield  {journal} {\bibinfo  {journal} {Quantum Inf. Process.}\ }\textbf {\bibinfo {volume} {1}},\ \bibinfo {pages} {5} (\bibinfo {year} {2002})}\BibitemShut {NoStop}%
\bibitem [{\citenamefont {Horodecki}\ \emph {et~al.}(2006)\citenamefont {Horodecki}, \citenamefont {Oppenheim},\ and\ \citenamefont {Winter}}]{horodecki2006a}%
  \BibitemOpen
  \bibfield  {author} {\bibinfo {author} {\bibfnamefont {M.}~\bibnamefont {Horodecki}}, \bibinfo {author} {\bibfnamefont {J.}~\bibnamefont {Oppenheim}},\ and\ \bibinfo {author} {\bibfnamefont {A.}~\bibnamefont {Winter}},\ }\bibfield  {title} {\bibinfo {title} {Quantum {{State Merging}} and {{Negative Information}}},\ }\href {https://doi.org/10.1007/s00220-006-0118-x} {\bibfield  {journal} {\bibinfo  {journal} {Commun. Math. Phys.}\ }\textbf {\bibinfo {volume} {269}},\ \bibinfo {pages} {107} (\bibinfo {year} {2006})}\BibitemShut {NoStop}%
\bibitem [{\citenamefont {Schumacher}(1996)}]{schumacher1996b}%
  \BibitemOpen
  \bibfield  {author} {\bibinfo {author} {\bibfnamefont {B.}~\bibnamefont {Schumacher}},\ }\bibfield  {title} {\bibinfo {title} {Sending entanglement through noisy quantum channels},\ }\href {https://doi.org/10.1103/PhysRevA.54.2614} {\bibfield  {journal} {\bibinfo  {journal} {Phys. Rev. A}\ }\textbf {\bibinfo {volume} {54}},\ \bibinfo {pages} {2614} (\bibinfo {year} {1996})}\BibitemShut {NoStop}%
\bibitem [{\citenamefont {Wilde}(2013)}]{wilde2013}%
  \BibitemOpen
  \bibfield  {author} {\bibinfo {author} {\bibfnamefont {M.~M.}\ \bibnamefont {Wilde}},\ }\href {https://doi.org/10.1017/CBO9781139525343} {\emph {\bibinfo {title} {Quantum information theory}}}\ (\bibinfo  {publisher} {Cambridge university press},\ \bibinfo {year} {2013})\BibitemShut {NoStop}%
\bibitem [{\citenamefont {Gottesman}(2009)}]{gottesman2009}%
  \BibitemOpen
  \bibfield  {author} {\bibinfo {author} {\bibfnamefont {D.}~\bibnamefont {Gottesman}},\ }\href {https://arxiv.org/abs/0904.2557} {\bibinfo {title} {An introduction to quantum error correction and fault-tolerant quantum computation}} (\bibinfo {year} {2009}),\ \Eprint {https://arxiv.org/abs/0904.2557} {arXiv:0904.2557 [quant-ph]} \BibitemShut {NoStop}%
\bibitem [{\citenamefont {Dennis}\ \emph {et~al.}(2002)\citenamefont {Dennis}, \citenamefont {Kitaev}, \citenamefont {Landahl},\ and\ \citenamefont {Preskill}}]{dennis2002}%
  \BibitemOpen
  \bibfield  {author} {\bibinfo {author} {\bibfnamefont {E.}~\bibnamefont {Dennis}}, \bibinfo {author} {\bibfnamefont {A.}~\bibnamefont {Kitaev}}, \bibinfo {author} {\bibfnamefont {A.}~\bibnamefont {Landahl}},\ and\ \bibinfo {author} {\bibfnamefont {J.}~\bibnamefont {Preskill}},\ }\bibfield  {title} {\bibinfo {title} {Topological quantum memory},\ }\href {https://doi.org/10.1063/1.1499754} {\bibfield  {journal} {\bibinfo  {journal} {J. Math. Phys.}\ }\textbf {\bibinfo {volume} {43}},\ \bibinfo {pages} {4452} (\bibinfo {year} {2002})}\BibitemShut {NoStop}%
\bibitem [{\citenamefont {Wang}\ \emph {et~al.}(2003)\citenamefont {Wang}, \citenamefont {Harrington},\ and\ \citenamefont {Preskill}}]{wang2003}%
  \BibitemOpen
  \bibfield  {author} {\bibinfo {author} {\bibfnamefont {C.}~\bibnamefont {Wang}}, \bibinfo {author} {\bibfnamefont {J.}~\bibnamefont {Harrington}},\ and\ \bibinfo {author} {\bibfnamefont {J.}~\bibnamefont {Preskill}},\ }\bibfield  {title} {\bibinfo {title} {Confinement-{{Higgs}} transition in a disordered gauge theory and the accuracy threshold for quantum memory},\ }\href {https://doi.org/https://doi.org/10.1016/S0003-4916(02)00019-2} {\bibfield  {journal} {\bibinfo  {journal} {Ann. Phys.}\ }\textbf {\bibinfo {volume} {303}},\ \bibinfo {pages} {31} (\bibinfo {year} {2003})}\BibitemShut {NoStop}%
\bibitem [{\citenamefont {Kitaev}(1997)}]{kitaev1997}%
  \BibitemOpen
  \bibfield  {author} {\bibinfo {author} {\bibfnamefont {A.~{\relax Yu}.}\ \bibnamefont {Kitaev}},\ }\bibfield  {title} {\bibinfo {title} {Quantum {{Error Correction}} with {{Imperfect Gates}}},\ }in\ \href {https://doi.org/10.1007/978-1-4615-5923-8_19} {\emph {\bibinfo {booktitle} {Quantum {{Communication}}, {{Computing}}, and {{Measurement}}}}},\ \bibinfo {editor} {edited by\ \bibinfo {editor} {\bibfnamefont {O.}~\bibnamefont {Hirota}}, \bibinfo {editor} {\bibfnamefont {A.~S.}\ \bibnamefont {Holevo}},\ and\ \bibinfo {editor} {\bibfnamefont {C.~M.}\ \bibnamefont {Caves}}}\ (\bibinfo  {publisher} {Springer US},\ \bibinfo {address} {Boston, MA},\ \bibinfo {year} {1997})\ pp.\ \bibinfo {pages} {181--188}\BibitemShut {NoStop}%
\bibitem [{\citenamefont {Terhal}(2015)}]{terhal2015}%
  \BibitemOpen
  \bibfield  {author} {\bibinfo {author} {\bibfnamefont {B.~M.}\ \bibnamefont {Terhal}},\ }\bibfield  {title} {\bibinfo {title} {Quantum error correction for quantum memories},\ }\href {https://journals.aps.org/rmp/abstract/10.1103/RevModPhys.87.307} {\bibfield  {journal} {\bibinfo  {journal} {Rev. Mod. Phys.}\ }\textbf {\bibinfo {volume} {87}},\ \bibinfo {pages} {307} (\bibinfo {year} {2015})}\BibitemShut {NoStop}%
\bibitem [{\citenamefont {Kitaev}(2003)}]{kitaev2003a}%
  \BibitemOpen
  \bibfield  {author} {\bibinfo {author} {\bibfnamefont {A.~Y.}\ \bibnamefont {Kitaev}},\ }\bibfield  {title} {\bibinfo {title} {Fault-tolerant quantum computation by anyons},\ }\href {https://doi.org/10.1016/S0003-4916(02)00018-0} {\bibfield  {journal} {\bibinfo  {journal} {Ann. Phys.}\ }\textbf {\bibinfo {volume} {303}},\ \bibinfo {pages} {2} (\bibinfo {year} {2003})}\BibitemShut {NoStop}%
\bibitem [{\citenamefont {Sekino}\ and\ \citenamefont {Susskind}(2008)}]{sekino2008}%
  \BibitemOpen
  \bibfield  {author} {\bibinfo {author} {\bibfnamefont {Y.}~\bibnamefont {Sekino}}\ and\ \bibinfo {author} {\bibfnamefont {L.}~\bibnamefont {Susskind}},\ }\bibfield  {title} {\bibinfo {title} {Fast scramblers},\ }\href {https://iopscience.iop.org/article/10.1088/1126-6708/2008/10/065} {\bibfield  {journal} {\bibinfo  {journal} {J. High Energy Phys.}\ }\textbf {\bibinfo {volume} {2008}}\bibinfo  {number} { (10)},\ \bibinfo {pages} {065}}\BibitemShut {NoStop}%
\bibitem [{\citenamefont {Mi}\ \emph {et~al.}(2021)\citenamefont {Mi}, \citenamefont {Roushan}, \citenamefont {Quintana}, \citenamefont {Mandra}, \citenamefont {Marshall}, \citenamefont {Neill}, \citenamefont {Arute}, \citenamefont {Arya}, \citenamefont {Atalaya}, \citenamefont {Babbush} \emph {et~al.}}]{mi2021}%
  \BibitemOpen
\bibfield  {number} {  }\bibfield  {author} {\bibinfo {author} {\bibfnamefont {X.}~\bibnamefont {Mi}}, \bibinfo {author} {\bibfnamefont {P.}~\bibnamefont {Roushan}}, \bibinfo {author} {\bibfnamefont {C.}~\bibnamefont {Quintana}}, \bibinfo {author} {\bibfnamefont {S.}~\bibnamefont {Mandra}}, \bibinfo {author} {\bibfnamefont {J.}~\bibnamefont {Marshall}}, \bibinfo {author} {\bibfnamefont {C.}~\bibnamefont {Neill}}, \bibinfo {author} {\bibfnamefont {F.}~\bibnamefont {Arute}}, \bibinfo {author} {\bibfnamefont {K.}~\bibnamefont {Arya}}, \bibinfo {author} {\bibfnamefont {J.}~\bibnamefont {Atalaya}}, \bibinfo {author} {\bibfnamefont {R.}~\bibnamefont {Babbush}}, \emph {et~al.},\ }\bibfield  {title} {\bibinfo {title} {Information scrambling in quantum circuits},\ }\href {https://doi.org/10.1126/science.abg5029} {\bibfield  {journal} {\bibinfo  {journal} {Science}\ }\textbf {\bibinfo {volume} {374}},\ \bibinfo {pages} {1479} (\bibinfo {year} {2021})}\BibitemShut {NoStop}%
\bibitem [{\citenamefont {Landsman}\ \emph {et~al.}(2019)\citenamefont {Landsman}, \citenamefont {Figgatt}, \citenamefont {Schuster}, \citenamefont {Linke}, \citenamefont {Yoshida}, \citenamefont {Yao},\ and\ \citenamefont {Monroe}}]{landsman2019}%
  \BibitemOpen
  \bibfield  {author} {\bibinfo {author} {\bibfnamefont {K.~A.}\ \bibnamefont {Landsman}}, \bibinfo {author} {\bibfnamefont {C.}~\bibnamefont {Figgatt}}, \bibinfo {author} {\bibfnamefont {T.}~\bibnamefont {Schuster}}, \bibinfo {author} {\bibfnamefont {N.~M.}\ \bibnamefont {Linke}}, \bibinfo {author} {\bibfnamefont {B.}~\bibnamefont {Yoshida}}, \bibinfo {author} {\bibfnamefont {N.~Y.}\ \bibnamefont {Yao}},\ and\ \bibinfo {author} {\bibfnamefont {C.}~\bibnamefont {Monroe}},\ }\bibfield  {title} {\bibinfo {title} {Verified quantum information scrambling},\ }\href {https://www.nature.com/articles/s41586-019-0952-6} {\bibfield  {journal} {\bibinfo  {journal} {Nature}\ }\textbf {\bibinfo {volume} {567}},\ \bibinfo {pages} {61} (\bibinfo {year} {2019})}\BibitemShut {NoStop}%
\bibitem [{\citenamefont {Li}\ \emph {et~al.}(2019)\citenamefont {Li}, \citenamefont {Chen},\ and\ \citenamefont {Fisher}}]{li2019}%
  \BibitemOpen
  \bibfield  {author} {\bibinfo {author} {\bibfnamefont {Y.}~\bibnamefont {Li}}, \bibinfo {author} {\bibfnamefont {X.}~\bibnamefont {Chen}},\ and\ \bibinfo {author} {\bibfnamefont {M.~P.~A.}\ \bibnamefont {Fisher}},\ }\bibfield  {title} {\bibinfo {title} {Measurement-driven entanglement transition in hybrid quantum circuits},\ }\href {https://doi.org/10.1103/PhysRevB.100.134306} {\bibfield  {journal} {\bibinfo  {journal} {Phys. Rev. B}\ }\textbf {\bibinfo {volume} {100}},\ \bibinfo {pages} {134306} (\bibinfo {year} {2019})}\BibitemShut {NoStop}%
\bibitem [{\citenamefont {Chan}\ \emph {et~al.}(2019)\citenamefont {Chan}, \citenamefont {Nandkishore}, \citenamefont {Pretko},\ and\ \citenamefont {Smith}}]{chan2019}%
  \BibitemOpen
  \bibfield  {author} {\bibinfo {author} {\bibfnamefont {A.}~\bibnamefont {Chan}}, \bibinfo {author} {\bibfnamefont {R.~M.}\ \bibnamefont {Nandkishore}}, \bibinfo {author} {\bibfnamefont {M.}~\bibnamefont {Pretko}},\ and\ \bibinfo {author} {\bibfnamefont {G.}~\bibnamefont {Smith}},\ }\bibfield  {title} {\bibinfo {title} {Unitary-projective entanglement dynamics},\ }\href {https://doi.org/10.1103/PhysRevB.99.224307} {\bibfield  {journal} {\bibinfo  {journal} {Phys. Rev. B}\ }\textbf {\bibinfo {volume} {99}},\ \bibinfo {pages} {224307} (\bibinfo {year} {2019})}\BibitemShut {NoStop}%
\bibitem [{\citenamefont {Bao}\ \emph {et~al.}(2024)\citenamefont {Bao}, \citenamefont {Block},\ and\ \citenamefont {Altman}}]{bao2024}%
  \BibitemOpen
  \bibfield  {author} {\bibinfo {author} {\bibfnamefont {Y.}~\bibnamefont {Bao}}, \bibinfo {author} {\bibfnamefont {M.}~\bibnamefont {Block}},\ and\ \bibinfo {author} {\bibfnamefont {E.}~\bibnamefont {Altman}},\ }\bibfield  {title} {\bibinfo {title} {Finite-{{Time Teleportation Phase Transition}} in {{Random Quantum Circuits}}},\ }\href {https://doi.org/10.1103/PhysRevLett.132.030401} {\bibfield  {journal} {\bibinfo  {journal} {Phys. Rev. Lett.}\ }\textbf {\bibinfo {volume} {132}},\ \bibinfo {pages} {030401} (\bibinfo {year} {2024})}\BibitemShut {NoStop}%
\bibitem [{\citenamefont {Lee}\ \emph {et~al.}(2022)\citenamefont {Lee}, \citenamefont {Ji}, \citenamefont {Bi},\ and\ \citenamefont {Fisher}}]{lee2022}%
  \BibitemOpen
  \bibfield  {author} {\bibinfo {author} {\bibfnamefont {J.~Y.}\ \bibnamefont {Lee}}, \bibinfo {author} {\bibfnamefont {W.}~\bibnamefont {Ji}}, \bibinfo {author} {\bibfnamefont {Z.}~\bibnamefont {Bi}},\ and\ \bibinfo {author} {\bibfnamefont {M.~P.~A.}\ \bibnamefont {Fisher}},\ }\href {https://arxiv.org/abs/2208.11699} {\bibinfo {title} {Decoding measurement-prepared quantum phases and transitions: from ising model to gauge theory, and beyond}} (\bibinfo {year} {2022}),\ \Eprint {https://arxiv.org/abs/2208.11699} {arXiv:2208.11699 [cond-mat.str-el]} \BibitemShut {NoStop}%
\bibitem [{\citenamefont {Li}\ \emph {et~al.}(2023{\natexlab{a}})\citenamefont {Li}, \citenamefont {Vijay},\ and\ \citenamefont {Fisher}}]{li2023a}%
  \BibitemOpen
  \bibfield  {author} {\bibinfo {author} {\bibfnamefont {Y.}~\bibnamefont {Li}}, \bibinfo {author} {\bibfnamefont {S.}~\bibnamefont {Vijay}},\ and\ \bibinfo {author} {\bibfnamefont {M.~P.}\ \bibnamefont {Fisher}},\ }\bibfield  {title} {\bibinfo {title} {Entanglement domain walls in monitored quantum circuits and the directed polymer in a random environment},\ }\href {https://doi.org/10.1103/PRXQuantum.4.010331} {\bibfield  {journal} {\bibinfo  {journal} {PRX Quantum}\ }\textbf {\bibinfo {volume} {4}},\ \bibinfo {pages} {010331} (\bibinfo {year} {2023}{\natexlab{a}})}\BibitemShut {NoStop}%
\bibitem [{\citenamefont {Nahum}\ and\ \citenamefont {Skinner}(2020)}]{nahum2020}%
  \BibitemOpen
  \bibfield  {author} {\bibinfo {author} {\bibfnamefont {A.}~\bibnamefont {Nahum}}\ and\ \bibinfo {author} {\bibfnamefont {B.}~\bibnamefont {Skinner}},\ }\bibfield  {title} {\bibinfo {title} {Entanglement and dynamics of diffusion-annihilation processes with majorana defects},\ }\href {https://doi.org/10.1103/PhysRevResearch.2.023288} {\bibfield  {journal} {\bibinfo  {journal} {Phys. Rev. Research}\ }\textbf {\bibinfo {volume} {2}},\ \bibinfo {pages} {023288} (\bibinfo {year} {2020})}\BibitemShut {NoStop}%
\bibitem [{\citenamefont {Nahum}\ \emph {et~al.}(2018)\citenamefont {Nahum}, \citenamefont {Vijay},\ and\ \citenamefont {Haah}}]{nahum2018}%
  \BibitemOpen
  \bibfield  {author} {\bibinfo {author} {\bibfnamefont {A.}~\bibnamefont {Nahum}}, \bibinfo {author} {\bibfnamefont {S.}~\bibnamefont {Vijay}},\ and\ \bibinfo {author} {\bibfnamefont {J.}~\bibnamefont {Haah}},\ }\bibfield  {title} {\bibinfo {title} {Operator spreading in random unitary circuits},\ }\href {https://doi.org/10.1103/PhysRevX.8.021014} {\bibfield  {journal} {\bibinfo  {journal} {Phys. Rev. X}\ }\textbf {\bibinfo {volume} {8}},\ \bibinfo {pages} {021014} (\bibinfo {year} {2018})}\BibitemShut {NoStop}%
\bibitem [{\citenamefont {Nahum}\ \emph {et~al.}(2017)\citenamefont {Nahum}, \citenamefont {Ruhman}, \citenamefont {Vijay},\ and\ \citenamefont {Haah}}]{nahum2019}%
  \BibitemOpen
  \bibfield  {author} {\bibinfo {author} {\bibfnamefont {A.}~\bibnamefont {Nahum}}, \bibinfo {author} {\bibfnamefont {J.}~\bibnamefont {Ruhman}}, \bibinfo {author} {\bibfnamefont {S.}~\bibnamefont {Vijay}},\ and\ \bibinfo {author} {\bibfnamefont {J.}~\bibnamefont {Haah}},\ }\bibfield  {title} {\bibinfo {title} {Quantum entanglement growth under random unitary dynamics},\ }\href {https://doi.org/10.1103/PhysRevX.7.031016} {\bibfield  {journal} {\bibinfo  {journal} {Phys. Rev. X}\ }\textbf {\bibinfo {volume} {7}},\ \bibinfo {pages} {031016} (\bibinfo {year} {2017})}\BibitemShut {NoStop}%
\bibitem [{\citenamefont {Sharma}\ \emph {et~al.}(2022)\citenamefont {Sharma}, \citenamefont {Turkeshi}, \citenamefont {Fazio},\ and\ \citenamefont {Dalmonte}}]{sharma2022}%
  \BibitemOpen
  \bibfield  {author} {\bibinfo {author} {\bibfnamefont {S.}~\bibnamefont {Sharma}}, \bibinfo {author} {\bibfnamefont {X.}~\bibnamefont {Turkeshi}}, \bibinfo {author} {\bibfnamefont {R.}~\bibnamefont {Fazio}},\ and\ \bibinfo {author} {\bibfnamefont {M.}~\bibnamefont {Dalmonte}},\ }\bibfield  {title} {\bibinfo {title} {Measurement-induced criticality in extended and long-range unitary circuits},\ }\href {https://doi.org/10.21468/SciPostPhysCore.5.2.023} {\bibfield  {journal} {\bibinfo  {journal} {SciPost Phys. Core}\ }\textbf {\bibinfo {volume} {5}},\ \bibinfo {pages} {023} (\bibinfo {year} {2022})}\BibitemShut {NoStop}%
\bibitem [{\citenamefont {Sang}\ \emph {et~al.}(2021)\citenamefont {Sang}, \citenamefont {Li}, \citenamefont {Zhou}, \citenamefont {Chen}, \citenamefont {Hsieh},\ and\ \citenamefont {Fisher}}]{sang2021a}%
  \BibitemOpen
  \bibfield  {author} {\bibinfo {author} {\bibfnamefont {S.}~\bibnamefont {Sang}}, \bibinfo {author} {\bibfnamefont {Y.}~\bibnamefont {Li}}, \bibinfo {author} {\bibfnamefont {T.}~\bibnamefont {Zhou}}, \bibinfo {author} {\bibfnamefont {X.}~\bibnamefont {Chen}}, \bibinfo {author} {\bibfnamefont {T.~H.}\ \bibnamefont {Hsieh}},\ and\ \bibinfo {author} {\bibfnamefont {M.~P.}\ \bibnamefont {Fisher}},\ }\bibfield  {title} {\bibinfo {title} {Entanglement negativity at measurement-induced criticality},\ }\href {https://doi.org/10.1103/PRXQuantum.2.030313} {\bibfield  {journal} {\bibinfo  {journal} {PRX Quantum}\ }\textbf {\bibinfo {volume} {2}},\ \bibinfo {pages} {030313} (\bibinfo {year} {2021})}\BibitemShut {NoStop}%
\bibitem [{\citenamefont {Skinner}\ \emph {et~al.}(2019)\citenamefont {Skinner}, \citenamefont {Ruhman},\ and\ \citenamefont {Nahum}}]{skinner2019}%
  \BibitemOpen
  \bibfield  {author} {\bibinfo {author} {\bibfnamefont {B.}~\bibnamefont {Skinner}}, \bibinfo {author} {\bibfnamefont {J.}~\bibnamefont {Ruhman}},\ and\ \bibinfo {author} {\bibfnamefont {A.}~\bibnamefont {Nahum}},\ }\bibfield  {title} {\bibinfo {title} {Measurement-induced phase transitions in the dynamics of entanglement},\ }\href {https://doi.org/10.1103/PhysRevX.9.031009} {\bibfield  {journal} {\bibinfo  {journal} {Phys. Rev. X}\ }\textbf {\bibinfo {volume} {9}},\ \bibinfo {pages} {031009} (\bibinfo {year} {2019})}\BibitemShut {NoStop}%
\bibitem [{\citenamefont {Szyniszewski}\ \emph {et~al.}(2019)\citenamefont {Szyniszewski}, \citenamefont {Romito},\ and\ \citenamefont {Schomerus}}]{szyniszewski2019}%
  \BibitemOpen
  \bibfield  {author} {\bibinfo {author} {\bibfnamefont {M.}~\bibnamefont {Szyniszewski}}, \bibinfo {author} {\bibfnamefont {A.}~\bibnamefont {Romito}},\ and\ \bibinfo {author} {\bibfnamefont {H.}~\bibnamefont {Schomerus}},\ }\bibfield  {title} {\bibinfo {title} {Entanglement transition from variable-strength weak measurements},\ }\href {https://doi.org/10.1103/PhysRevB.100.064204} {\bibfield  {journal} {\bibinfo  {journal} {Phys. Rev. B}\ }\textbf {\bibinfo {volume} {100}},\ \bibinfo {pages} {064204} (\bibinfo {year} {2019})}\BibitemShut {NoStop}%
\bibitem [{\citenamefont {Vasseur}\ \emph {et~al.}(2019)\citenamefont {Vasseur}, \citenamefont {Potter}, \citenamefont {You},\ and\ \citenamefont {Ludwig}}]{vasseur2019}%
  \BibitemOpen
  \bibfield  {author} {\bibinfo {author} {\bibfnamefont {R.}~\bibnamefont {Vasseur}}, \bibinfo {author} {\bibfnamefont {A.~C.}\ \bibnamefont {Potter}}, \bibinfo {author} {\bibfnamefont {Y.-Z.}\ \bibnamefont {You}},\ and\ \bibinfo {author} {\bibfnamefont {A.~W.~W.}\ \bibnamefont {Ludwig}},\ }\bibfield  {title} {\bibinfo {title} {Entanglement transitions from holographic random tensor networks},\ }\href {https://doi.org/10.1103/PhysRevB.100.134203} {\bibfield  {journal} {\bibinfo  {journal} {Phys. Rev. B}\ }\textbf {\bibinfo {volume} {100}},\ \bibinfo {pages} {134203} (\bibinfo {year} {2019})}\BibitemShut {NoStop}%
\bibitem [{\citenamefont {Zabalo}\ \emph {et~al.}(2020)\citenamefont {Zabalo}, \citenamefont {Gullans}, \citenamefont {Wilson}, \citenamefont {Gopalakrishnan}, \citenamefont {Huse},\ and\ \citenamefont {Pixley}}]{zabalo2020}%
  \BibitemOpen
  \bibfield  {author} {\bibinfo {author} {\bibfnamefont {A.}~\bibnamefont {Zabalo}}, \bibinfo {author} {\bibfnamefont {M.~J.}\ \bibnamefont {Gullans}}, \bibinfo {author} {\bibfnamefont {J.~H.}\ \bibnamefont {Wilson}}, \bibinfo {author} {\bibfnamefont {S.}~\bibnamefont {Gopalakrishnan}}, \bibinfo {author} {\bibfnamefont {D.~A.}\ \bibnamefont {Huse}},\ and\ \bibinfo {author} {\bibfnamefont {J.~H.}\ \bibnamefont {Pixley}},\ }\bibfield  {title} {\bibinfo {title} {Critical properties of the measurement-induced transition in random quantum circuits},\ }\href {https://doi.org/10.1103/PhysRevB.101.060301} {\bibfield  {journal} {\bibinfo  {journal} {Phys. Rev. B}\ }\textbf {\bibinfo {volume} {101}},\ \bibinfo {pages} {060301} (\bibinfo {year} {2020})}\BibitemShut {NoStop}%
\bibitem [{\citenamefont {Alberton}\ \emph {et~al.}(2021)\citenamefont {Alberton}, \citenamefont {Buchhold},\ and\ \citenamefont {Diehl}}]{alberton2021}%
  \BibitemOpen
  \bibfield  {author} {\bibinfo {author} {\bibfnamefont {O.}~\bibnamefont {Alberton}}, \bibinfo {author} {\bibfnamefont {M.}~\bibnamefont {Buchhold}},\ and\ \bibinfo {author} {\bibfnamefont {S.}~\bibnamefont {Diehl}},\ }\bibfield  {title} {\bibinfo {title} {Entanglement {{Transition}} in a {{Monitored Free-Fermion Chain}}: {{From Extended Criticality}} to {{Area Law}}},\ }\href {https://doi.org/10.1103/PhysRevLett.126.170602} {\bibfield  {journal} {\bibinfo  {journal} {Phys. Rev. Lett.}\ }\textbf {\bibinfo {volume} {126}},\ \bibinfo {pages} {170602} (\bibinfo {year} {2021})}\BibitemShut {NoStop}%
\bibitem [{\citenamefont {Fidkowski}\ \emph {et~al.}(2021)\citenamefont {Fidkowski}, \citenamefont {Haah},\ and\ \citenamefont {Hastings}}]{fidkowski2021}%
  \BibitemOpen
  \bibfield  {author} {\bibinfo {author} {\bibfnamefont {L.}~\bibnamefont {Fidkowski}}, \bibinfo {author} {\bibfnamefont {J.}~\bibnamefont {Haah}},\ and\ \bibinfo {author} {\bibfnamefont {M.~B.}\ \bibnamefont {Hastings}},\ }\bibfield  {title} {\bibinfo {title} {How dynamical quantum memories forget},\ }\href {https://quantum-journal.org/papers/q-2021-01-17-382/} {\bibfield  {journal} {\bibinfo  {journal} {Quantum}\ }\textbf {\bibinfo {volume} {5}},\ \bibinfo {pages} {382} (\bibinfo {year} {2021})}\BibitemShut {NoStop}%
\bibitem [{\citenamefont {Fisher}\ \emph {et~al.}(2023)\citenamefont {Fisher}, \citenamefont {Khemani}, \citenamefont {Nahum},\ and\ \citenamefont {Vijay}}]{fisher2023}%
  \BibitemOpen
  \bibfield  {author} {\bibinfo {author} {\bibfnamefont {M.~P.~A.}\ \bibnamefont {Fisher}}, \bibinfo {author} {\bibfnamefont {V.}~\bibnamefont {Khemani}}, \bibinfo {author} {\bibfnamefont {A.}~\bibnamefont {Nahum}},\ and\ \bibinfo {author} {\bibfnamefont {S.}~\bibnamefont {Vijay}},\ }\bibfield  {title} {\bibinfo {title} {Random {{Quantum Circuits}}},\ }\href {https://doi.org/10.1146/annurev-conmatphys-031720-030658} {\bibfield  {journal} {\bibinfo  {journal} {Annu. Rev. Condens. Matter Phys.}\ }\textbf {\bibinfo {volume} {14}},\ \bibinfo {pages} {335} (\bibinfo {year} {2023})}\BibitemShut {NoStop}%
\bibitem [{\citenamefont {Poboiko}\ \emph {et~al.}(2024)\citenamefont {Poboiko}, \citenamefont {Gornyi},\ and\ \citenamefont {Mirlin}}]{poboiko2024}%
  \BibitemOpen
  \bibfield  {author} {\bibinfo {author} {\bibfnamefont {I.}~\bibnamefont {Poboiko}}, \bibinfo {author} {\bibfnamefont {I.~V.}\ \bibnamefont {Gornyi}},\ and\ \bibinfo {author} {\bibfnamefont {A.~D.}\ \bibnamefont {Mirlin}},\ }\bibfield  {title} {\bibinfo {title} {Measurement-induced phase transition for free fermions above one dimension},\ }\href {https://doi.org/10.1103/PhysRevLett.132.110403} {\bibfield  {journal} {\bibinfo  {journal} {Phys. Rev. Lett.}\ }\textbf {\bibinfo {volume} {132}},\ \bibinfo {pages} {110403} (\bibinfo {year} {2024})}\BibitemShut {NoStop}%
\bibitem [{\citenamefont {Yu}\ and\ \citenamefont {Qi}(2022)}]{yu2022}%
  \BibitemOpen
  \bibfield  {author} {\bibinfo {author} {\bibfnamefont {X.}~\bibnamefont {Yu}}\ and\ \bibinfo {author} {\bibfnamefont {X.-L.}\ \bibnamefont {Qi}},\ }\href {https://arxiv.org/abs/2201.12704} {\bibinfo {title} {Measurement-induced entanglement phase transition in random bilocal circuits}} (\bibinfo {year} {2022}),\ \Eprint {https://arxiv.org/abs/2201.12704} {arXiv:2201.12704 [quant-ph]} \BibitemShut {NoStop}%
\bibitem [{\citenamefont {Choi}\ \emph {et~al.}(2020)\citenamefont {Choi}, \citenamefont {Bao}, \citenamefont {Qi},\ and\ \citenamefont {Altman}}]{choi2020}%
  \BibitemOpen
  \bibfield  {author} {\bibinfo {author} {\bibfnamefont {S.}~\bibnamefont {Choi}}, \bibinfo {author} {\bibfnamefont {Y.}~\bibnamefont {Bao}}, \bibinfo {author} {\bibfnamefont {X.-L.}\ \bibnamefont {Qi}},\ and\ \bibinfo {author} {\bibfnamefont {E.}~\bibnamefont {Altman}},\ }\bibfield  {title} {\bibinfo {title} {Quantum {{Error Correction}} in {{Scrambling Dynamics}} and {{Measurement-Induced Phase Transition}}},\ }\href {https://doi.org/10.1103/PhysRevLett.125.030505} {\bibfield  {journal} {\bibinfo  {journal} {Phys. Rev. Lett.}\ }\textbf {\bibinfo {volume} {125}},\ \bibinfo {pages} {030505} (\bibinfo {year} {2020})}\BibitemShut {NoStop}%
\bibitem [{\citenamefont {Gullans}\ and\ \citenamefont {Huse}(2020)}]{gullans2020a}%
  \BibitemOpen
  \bibfield  {author} {\bibinfo {author} {\bibfnamefont {M.~J.}\ \bibnamefont {Gullans}}\ and\ \bibinfo {author} {\bibfnamefont {D.~A.}\ \bibnamefont {Huse}},\ }\bibfield  {title} {\bibinfo {title} {Dynamical purification phase transition induced by quantum measurements},\ }\href {https://doi.org/10.1103/PhysRevX.10.041020} {\bibfield  {journal} {\bibinfo  {journal} {Phys. Rev. X}\ }\textbf {\bibinfo {volume} {10}},\ \bibinfo {pages} {041020} (\bibinfo {year} {2020})}\BibitemShut {NoStop}%
\bibitem [{\citenamefont {Dias}\ \emph {et~al.}(2023)\citenamefont {Dias}, \citenamefont {Perković}, \citenamefont {Haque}, \citenamefont {Ribeiro},\ and\ \citenamefont {McClarty}}]{dias2023}%
  \BibitemOpen
  \bibfield  {author} {\bibinfo {author} {\bibfnamefont {B.~C.}\ \bibnamefont {Dias}}, \bibinfo {author} {\bibfnamefont {D.}~\bibnamefont {Perković}}, \bibinfo {author} {\bibfnamefont {M.}~\bibnamefont {Haque}}, \bibinfo {author} {\bibfnamefont {P.}~\bibnamefont {Ribeiro}},\ and\ \bibinfo {author} {\bibfnamefont {P.~A.}\ \bibnamefont {McClarty}},\ }\bibfield  {title} {\bibinfo {title} {Quantum noise as a symmetry-breaking field},\ }\href {https://doi.org/10.1103/PhysRevB.108.L060302} {\bibfield  {journal} {\bibinfo  {journal} {Phys. Rev. B}\ }\textbf {\bibinfo {volume} {108}},\ \bibinfo {pages} {L060302} (\bibinfo {year} {2023})}\BibitemShut {NoStop}%
\bibitem [{\citenamefont {Liu}\ \emph {et~al.}(2023)\citenamefont {Liu}, \citenamefont {Li}, \citenamefont {Zhang}, \citenamefont {Jian},\ and\ \citenamefont {Yao}}]{liu2023e}%
  \BibitemOpen
  \bibfield  {author} {\bibinfo {author} {\bibfnamefont {S.}~\bibnamefont {Liu}}, \bibinfo {author} {\bibfnamefont {M.-R.}\ \bibnamefont {Li}}, \bibinfo {author} {\bibfnamefont {S.-X.}\ \bibnamefont {Zhang}}, \bibinfo {author} {\bibfnamefont {S.-K.}\ \bibnamefont {Jian}},\ and\ \bibinfo {author} {\bibfnamefont {H.}~\bibnamefont {Yao}},\ }\bibfield  {title} {\bibinfo {title} {Universal {{KPZ}} scaling in noisy hybrid quantum circuits},\ }\href {https://doi.org/10.1103/PhysRevB.107.L201113} {\bibfield  {journal} {\bibinfo  {journal} {Phys. Rev. B}\ }\textbf {\bibinfo {volume} {107}},\ \bibinfo {pages} {L201113} (\bibinfo {year} {2023})}\BibitemShut {NoStop}%
\bibitem [{\citenamefont {Liu}\ \emph {et~al.}(2024{\natexlab{a}})\citenamefont {Liu}, \citenamefont {Li}, \citenamefont {Zhang},\ and\ \citenamefont {Jian}}]{liu2024}%
  \BibitemOpen
  \bibfield  {author} {\bibinfo {author} {\bibfnamefont {S.}~\bibnamefont {Liu}}, \bibinfo {author} {\bibfnamefont {M.-R.}\ \bibnamefont {Li}}, \bibinfo {author} {\bibfnamefont {S.-X.}\ \bibnamefont {Zhang}},\ and\ \bibinfo {author} {\bibfnamefont {S.-K.}\ \bibnamefont {Jian}},\ }\bibfield  {title} {\bibinfo {title} {Entanglement structure and information protection in noisy hybrid quantum circuits},\ }\href {https://doi.org/10.1103/PhysRevLett.132.240402} {\bibfield  {journal} {\bibinfo  {journal} {Phys. Rev. Lett.}\ }\textbf {\bibinfo {volume} {132}},\ \bibinfo {pages} {240402} (\bibinfo {year} {2024}{\natexlab{a}})}\BibitemShut {NoStop}%
\bibitem [{\citenamefont {Liu}\ \emph {et~al.}(2024{\natexlab{b}})\citenamefont {Liu}, \citenamefont {Li}, \citenamefont {Zhang}, \citenamefont {Jian},\ and\ \citenamefont {Yao}}]{liu2024a}%
  \BibitemOpen
  \bibfield  {author} {\bibinfo {author} {\bibfnamefont {S.}~\bibnamefont {Liu}}, \bibinfo {author} {\bibfnamefont {M.-R.}\ \bibnamefont {Li}}, \bibinfo {author} {\bibfnamefont {S.-X.}\ \bibnamefont {Zhang}}, \bibinfo {author} {\bibfnamefont {S.-K.}\ \bibnamefont {Jian}},\ and\ \bibinfo {author} {\bibfnamefont {H.}~\bibnamefont {Yao}},\ }\bibfield  {title} {\bibinfo {title} {Noise-induced phase transitions in hybrid quantum circuits},\ }\href {https://doi.org/10.1103/PhysRevB.110.064323} {\bibfield  {journal} {\bibinfo  {journal} {Phys. Rev. B}\ }\textbf {\bibinfo {volume} {110}},\ \bibinfo {pages} {064323} (\bibinfo {year} {2024}{\natexlab{b}})}\BibitemShut {NoStop}%
\bibitem [{\citenamefont {Weinstein}\ \emph {et~al.}(2022)\citenamefont {Weinstein}, \citenamefont {Bao},\ and\ \citenamefont {Altman}}]{weinstein2022}%
  \BibitemOpen
  \bibfield  {author} {\bibinfo {author} {\bibfnamefont {Z.}~\bibnamefont {Weinstein}}, \bibinfo {author} {\bibfnamefont {Y.}~\bibnamefont {Bao}},\ and\ \bibinfo {author} {\bibfnamefont {E.}~\bibnamefont {Altman}},\ }\bibfield  {title} {\bibinfo {title} {Measurement-{{Induced Power-Law Negativity}} in an {{Open Monitored Quantum Circuit}}},\ }\href {https://doi.org/10.1103/PhysRevLett.129.080501} {\bibfield  {journal} {\bibinfo  {journal} {Phys. Rev. Lett.}\ }\textbf {\bibinfo {volume} {129}},\ \bibinfo {pages} {080501} (\bibinfo {year} {2022})}\BibitemShut {NoStop}%
\bibitem [{\citenamefont {Haas}\ \emph {et~al.}(2024)\citenamefont {Haas}, \citenamefont {Carisch},\ and\ \citenamefont {Zilberberg}}]{Haas2024}%
  \BibitemOpen
  \bibfield  {author} {\bibinfo {author} {\bibfnamefont {L.}~\bibnamefont {Haas}}, \bibinfo {author} {\bibfnamefont {C.}~\bibnamefont {Carisch}},\ and\ \bibinfo {author} {\bibfnamefont {O.}~\bibnamefont {Zilberberg}},\ }\href {https://arxiv.org/abs/2408.02810} {\bibinfo {title} {Scrambling-induced entanglement suppression in noisy quantum circuits}} (\bibinfo {year} {2024}),\ \Eprint {https://arxiv.org/abs/2408.02810} {arXiv:2408.02810 [quant-ph]} \BibitemShut {NoStop}%
\bibitem [{\citenamefont {Weinstein}\ \emph {et~al.}(2023)\citenamefont {Weinstein}, \citenamefont {Kelly}, \citenamefont {Marino},\ and\ \citenamefont {Altman}}]{Weinstein2024}%
  \BibitemOpen
  \bibfield  {author} {\bibinfo {author} {\bibfnamefont {Z.}~\bibnamefont {Weinstein}}, \bibinfo {author} {\bibfnamefont {S.~P.}\ \bibnamefont {Kelly}}, \bibinfo {author} {\bibfnamefont {J.}~\bibnamefont {Marino}},\ and\ \bibinfo {author} {\bibfnamefont {E.}~\bibnamefont {Altman}},\ }\bibfield  {title} {\bibinfo {title} {Scrambling transition in a radiative random unitary circuit},\ }\href {https://doi.org/10.1103/PhysRevLett.131.220404} {\bibfield  {journal} {\bibinfo  {journal} {Phys. Rev. Lett.}\ }\textbf {\bibinfo {volume} {131}},\ \bibinfo {pages} {220404} (\bibinfo {year} {2023})}\BibitemShut {NoStop}%
\bibitem [{\citenamefont {Lovas}\ \emph {et~al.}(2024)\citenamefont {Lovas}, \citenamefont {Agrawal},\ and\ \citenamefont {Vijay}}]{lovas2024}%
  \BibitemOpen
  \bibfield  {author} {\bibinfo {author} {\bibfnamefont {I.}~\bibnamefont {Lovas}}, \bibinfo {author} {\bibfnamefont {U.}~\bibnamefont {Agrawal}},\ and\ \bibinfo {author} {\bibfnamefont {S.}~\bibnamefont {Vijay}},\ }\bibfield  {title} {\bibinfo {title} {Quantum coding transitions in the presence of boundary dissipation},\ }\href {https://doi.org/10.1103/PRXQuantum.5.030327} {\bibfield  {journal} {\bibinfo  {journal} {PRX Quantum}\ }\textbf {\bibinfo {volume} {5}},\ \bibinfo {pages} {030327} (\bibinfo {year} {2024})}\BibitemShut {NoStop}%
\bibitem [{\citenamefont {Braun}\ \emph {et~al.}(2018)\citenamefont {Braun}, \citenamefont {Adesso}, \citenamefont {Benatti}, \citenamefont {Floreanini}, \citenamefont {Marzolino}, \citenamefont {Mitchell},\ and\ \citenamefont {Pirandola}}]{braun2018}%
  \BibitemOpen
  \bibfield  {author} {\bibinfo {author} {\bibfnamefont {D.}~\bibnamefont {Braun}}, \bibinfo {author} {\bibfnamefont {G.}~\bibnamefont {Adesso}}, \bibinfo {author} {\bibfnamefont {F.}~\bibnamefont {Benatti}}, \bibinfo {author} {\bibfnamefont {R.}~\bibnamefont {Floreanini}}, \bibinfo {author} {\bibfnamefont {U.}~\bibnamefont {Marzolino}}, \bibinfo {author} {\bibfnamefont {M.~W.}\ \bibnamefont {Mitchell}},\ and\ \bibinfo {author} {\bibfnamefont {S.}~\bibnamefont {Pirandola}},\ }\bibfield  {title} {\bibinfo {title} {Quantum-enhanced measurements without entanglement},\ }\href {https://doi.org/10.1103/RevModPhys.90.035006} {\bibfield  {journal} {\bibinfo  {journal} {Rev. Mod. Phys.}\ }\textbf {\bibinfo {volume} {90}},\ \bibinfo {pages} {035006} (\bibinfo {year} {2018})}\BibitemShut {NoStop}%
\bibitem [{\citenamefont {Aharonov}\ \emph {et~al.}(2022)\citenamefont {Aharonov}, \citenamefont {Cotler},\ and\ \citenamefont {Qi}}]{aharonov2022}%
  \BibitemOpen
  \bibfield  {author} {\bibinfo {author} {\bibfnamefont {D.}~\bibnamefont {Aharonov}}, \bibinfo {author} {\bibfnamefont {J.}~\bibnamefont {Cotler}},\ and\ \bibinfo {author} {\bibfnamefont {X.-L.}\ \bibnamefont {Qi}},\ }\bibfield  {title} {\bibinfo {title} {Quantum algorithmic measurement},\ }\href {https://doi.org/10.1038/s41467-021-27922-0} {\bibfield  {journal} {\bibinfo  {journal} {Nature Commun.}\ }\textbf {\bibinfo {volume} {13}},\ \bibinfo {pages} {887} (\bibinfo {year} {2022})}\BibitemShut {NoStop}%
\bibitem [{\citenamefont {Huang}\ \emph {et~al.}(2021)\citenamefont {Huang}, \citenamefont {Kueng},\ and\ \citenamefont {Preskill}}]{huang2021b}%
  \BibitemOpen
  \bibfield  {author} {\bibinfo {author} {\bibfnamefont {H.-Y.}\ \bibnamefont {Huang}}, \bibinfo {author} {\bibfnamefont {R.}~\bibnamefont {Kueng}},\ and\ \bibinfo {author} {\bibfnamefont {J.}~\bibnamefont {Preskill}},\ }\bibfield  {title} {\bibinfo {title} {Information-{{Theoretic Bounds}} on {{Quantum Advantage}} in {{Machine Learning}}},\ }\href {https://doi.org/10.1103/PhysRevLett.126.190505} {\bibfield  {journal} {\bibinfo  {journal} {Phys. Rev. Lett.}\ }\textbf {\bibinfo {volume} {126}},\ \bibinfo {pages} {190505} (\bibinfo {year} {2021})}\BibitemShut {NoStop}%
\bibitem [{\citenamefont {Zhao}\ \emph {et~al.}(2017)\citenamefont {Zhao}, \citenamefont {Dias}, \citenamefont {Haw}, \citenamefont {Bradshaw}, \citenamefont {Blandino}, \citenamefont {Symul}, \citenamefont {Ralph}, \citenamefont {Assad},\ and\ \citenamefont {Lam}}]{zhao2017}%
  \BibitemOpen
  \bibfield  {author} {\bibinfo {author} {\bibfnamefont {J.}~\bibnamefont {Zhao}}, \bibinfo {author} {\bibfnamefont {J.}~\bibnamefont {Dias}}, \bibinfo {author} {\bibfnamefont {J.~Y.}\ \bibnamefont {Haw}}, \bibinfo {author} {\bibfnamefont {M.}~\bibnamefont {Bradshaw}}, \bibinfo {author} {\bibfnamefont {R.}~\bibnamefont {Blandino}}, \bibinfo {author} {\bibfnamefont {T.}~\bibnamefont {Symul}}, \bibinfo {author} {\bibfnamefont {T.~C.}\ \bibnamefont {Ralph}}, \bibinfo {author} {\bibfnamefont {S.~M.}\ \bibnamefont {Assad}},\ and\ \bibinfo {author} {\bibfnamefont {P.~K.}\ \bibnamefont {Lam}},\ }\bibfield  {title} {\bibinfo {title} {Quantum enhancement of signal-to-noise ratio with a heralded linear amplifier},\ }\href {https://doi.org/10.1364/OPTICA.4.001421} {\bibfield  {journal} {\bibinfo  {journal} {Optica}\ }\textbf {\bibinfo {volume} {4}},\ \bibinfo {pages} {1421} (\bibinfo {year} {2017})}\BibitemShut {NoStop}%
\bibitem [{\citenamefont {Kelly}\ and\ \citenamefont {Marino}(2025{\natexlab{a}})}]{kelly2024a}%
  \BibitemOpen
  \bibfield  {author} {\bibinfo {author} {\bibfnamefont {S.~P.}\ \bibnamefont {Kelly}}\ and\ \bibinfo {author} {\bibfnamefont {J.}~\bibnamefont {Marino}},\ }\bibfield  {title} {\bibinfo {title} {Entanglement transitions induced by quantum-data collection},\ }\href {https://doi.org/10.1103/PhysRevA.111.L010402} {\bibfield  {journal} {\bibinfo  {journal} {Phys. Rev. A}\ }\textbf {\bibinfo {volume} {111}},\ \bibinfo {pages} {L010402} (\bibinfo {year} {2025}{\natexlab{a}})}\BibitemShut {NoStop}%
\bibitem [{\citenamefont {Kelly}\ and\ \citenamefont {Marino}(2025{\natexlab{b}})}]{kelly2024b}%
  \BibitemOpen
  \bibfield  {author} {\bibinfo {author} {\bibfnamefont {S.~P.}\ \bibnamefont {Kelly}}\ and\ \bibinfo {author} {\bibfnamefont {J.}~\bibnamefont {Marino}},\ }\bibfield  {title} {\bibinfo {title} {Generalizing measurement-induced phase transitions to information exchange symmetry breaking},\ }\href {https://doi.org/10.1103/PhysRevA.111.012425} {\bibfield  {journal} {\bibinfo  {journal} {Phys. Rev. A}\ }\textbf {\bibinfo {volume} {111}},\ \bibinfo {pages} {012425} (\bibinfo {year} {2025}{\natexlab{b}})}\BibitemShut {NoStop}%
\bibitem [{\citenamefont {Qian}\ and\ \citenamefont {Wang}(2025)}]{qian2024b}%
  \BibitemOpen
  \bibfield  {author} {\bibinfo {author} {\bibfnamefont {D.}~\bibnamefont {Qian}}\ and\ \bibinfo {author} {\bibfnamefont {J.}~\bibnamefont {Wang}},\ }\bibfield  {title} {\bibinfo {title} {Protect measurement-induced phase transition from noise},\ }\href {https://doi.org/10.1103/PhysRevLett.134.020403} {\bibfield  {journal} {\bibinfo  {journal} {Phys. Rev. Lett.}\ }\textbf {\bibinfo {volume} {134}},\ \bibinfo {pages} {020403} (\bibinfo {year} {2025})}\BibitemShut {NoStop}%
\bibitem [{\citenamefont {Li}\ \emph {et~al.}(2023{\natexlab{b}})\citenamefont {Li}, \citenamefont {Zou}, \citenamefont {Glorioso}, \citenamefont {Altman},\ and\ \citenamefont {Fisher}}]{li2023}%
  \BibitemOpen
  \bibfield  {author} {\bibinfo {author} {\bibfnamefont {Y.}~\bibnamefont {Li}}, \bibinfo {author} {\bibfnamefont {Y.}~\bibnamefont {Zou}}, \bibinfo {author} {\bibfnamefont {P.}~\bibnamefont {Glorioso}}, \bibinfo {author} {\bibfnamefont {E.}~\bibnamefont {Altman}},\ and\ \bibinfo {author} {\bibfnamefont {M.~P.~A.}\ \bibnamefont {Fisher}},\ }\bibfield  {title} {\bibinfo {title} {Cross entropy benchmark for measurement-induced phase transitions},\ }\href {https://doi.org/10.1103/PhysRevLett.130.220404} {\bibfield  {journal} {\bibinfo  {journal} {Phys. Rev. Lett.}\ }\textbf {\bibinfo {volume} {130}},\ \bibinfo {pages} {220404} (\bibinfo {year} {2023}{\natexlab{b}})}\BibitemShut {NoStop}%
\bibitem [{\citenamefont {Friedman}\ \emph {et~al.}(2023)\citenamefont {Friedman}, \citenamefont {Hart},\ and\ \citenamefont {Nandkishore}}]{friedman2023}%
  \BibitemOpen
  \bibfield  {author} {\bibinfo {author} {\bibfnamefont {A.~J.}\ \bibnamefont {Friedman}}, \bibinfo {author} {\bibfnamefont {O.}~\bibnamefont {Hart}},\ and\ \bibinfo {author} {\bibfnamefont {R.}~\bibnamefont {Nandkishore}},\ }\bibfield  {title} {\bibinfo {title} {Measurement-{{Induced Phases}} of {{Matter Require Feedback}}},\ }\href {https://doi.org/10.1103/PRXQuantum.4.040309} {\bibfield  {journal} {\bibinfo  {journal} {PRX Quantum}\ }\textbf {\bibinfo {volume} {4}},\ \bibinfo {pages} {040309} (\bibinfo {year} {2023})}\BibitemShut {NoStop}%
\bibitem [{\citenamefont {Bao}\ \emph {et~al.}(2020)\citenamefont {Bao}, \citenamefont {Choi},\ and\ \citenamefont {Altman}}]{bao2020}%
  \BibitemOpen
  \bibfield  {author} {\bibinfo {author} {\bibfnamefont {Y.}~\bibnamefont {Bao}}, \bibinfo {author} {\bibfnamefont {S.}~\bibnamefont {Choi}},\ and\ \bibinfo {author} {\bibfnamefont {E.}~\bibnamefont {Altman}},\ }\bibfield  {title} {\bibinfo {title} {Theory of the phase transition in random unitary circuits with measurements},\ }\href {https://doi.org/10.1103/PhysRevB.101.104301} {\bibfield  {journal} {\bibinfo  {journal} {Phys. Rev. B}\ }\textbf {\bibinfo {volume} {101}},\ \bibinfo {pages} {104301} (\bibinfo {year} {2020})}\BibitemShut {NoStop}%
\bibitem [{\citenamefont {Jian}\ \emph {et~al.}(2020)\citenamefont {Jian}, \citenamefont {You}, \citenamefont {Vasseur},\ and\ \citenamefont {Ludwig}}]{jian2020a}%
  \BibitemOpen
  \bibfield  {author} {\bibinfo {author} {\bibfnamefont {C.-M.}\ \bibnamefont {Jian}}, \bibinfo {author} {\bibfnamefont {Y.-Z.}\ \bibnamefont {You}}, \bibinfo {author} {\bibfnamefont {R.}~\bibnamefont {Vasseur}},\ and\ \bibinfo {author} {\bibfnamefont {A.~W.~W.}\ \bibnamefont {Ludwig}},\ }\bibfield  {title} {\bibinfo {title} {Measurement-induced criticality in random quantum circuits},\ }\href {https://doi.org/10.1103/PhysRevB.101.104302} {\bibfield  {journal} {\bibinfo  {journal} {Phys. Rev. B}\ }\textbf {\bibinfo {volume} {101}},\ \bibinfo {pages} {104302} (\bibinfo {year} {2020})}\BibitemShut {NoStop}%
\bibitem [{\citenamefont {Zhou}\ and\ \citenamefont {Nahum}(2019)}]{zhou2019}%
  \BibitemOpen
  \bibfield  {author} {\bibinfo {author} {\bibfnamefont {T.}~\bibnamefont {Zhou}}\ and\ \bibinfo {author} {\bibfnamefont {A.}~\bibnamefont {Nahum}},\ }\bibfield  {title} {\bibinfo {title} {Emergent statistical mechanics of entanglement in random unitary circuits},\ }\href {https://doi.org/10.1103/PhysRevB.99.174205} {\bibfield  {journal} {\bibinfo  {journal} {Phys. Rev. B}\ }\textbf {\bibinfo {volume} {99}},\ \bibinfo {pages} {174205} (\bibinfo {year} {2019})}\BibitemShut {NoStop}%
\bibitem [{\citenamefont {Li}\ \emph {et~al.}(2021)\citenamefont {Li}, \citenamefont {Chen}, \citenamefont {Ludwig},\ and\ \citenamefont {Fisher}}]{li2021a}%
  \BibitemOpen
  \bibfield  {author} {\bibinfo {author} {\bibfnamefont {Y.}~\bibnamefont {Li}}, \bibinfo {author} {\bibfnamefont {X.}~\bibnamefont {Chen}}, \bibinfo {author} {\bibfnamefont {A.~W.~W.}\ \bibnamefont {Ludwig}},\ and\ \bibinfo {author} {\bibfnamefont {M.~P.~A.}\ \bibnamefont {Fisher}},\ }\bibfield  {title} {\bibinfo {title} {Conformal invariance and quantum nonlocality in critical hybrid circuits},\ }\href {https://doi.org/10.1103/PhysRevB.104.104305} {\bibfield  {journal} {\bibinfo  {journal} {Phys. Rev. B}\ }\textbf {\bibinfo {volume} {104}},\ \bibinfo {pages} {104305} (\bibinfo {year} {2021})}\BibitemShut {NoStop}%
\bibitem [{sup()}]{supp}%
  \BibitemOpen
  \href@noop {} {\bibinfo {title} {See {{Supplemental Material}} for technical details on analytic mapping, resource efficient protocol, and numerics, which includes {Ref}.~\cite{collins2006}.}}\BibitemShut {Stop}%
\bibitem [{\citenamefont {Aaronson}\ and\ \citenamefont {Gottesman}(2004)}]{aaronson2004}%
  \BibitemOpen
  \bibfield  {author} {\bibinfo {author} {\bibfnamefont {S.}~\bibnamefont {Aaronson}}\ and\ \bibinfo {author} {\bibfnamefont {D.}~\bibnamefont {Gottesman}},\ }\bibfield  {title} {\bibinfo {title} {Improved simulation of stabilizer circuits},\ }\href {https://doi.org/10.1103/PhysRevA.70.052328} {\bibfield  {journal} {\bibinfo  {journal} {Phys. Rev. A}\ }\textbf {\bibinfo {volume} {70}},\ \bibinfo {pages} {052328} (\bibinfo {year} {2004})}\BibitemShut {NoStop}%
\bibitem [{\citenamefont {Gottesman}(1998)}]{gottesman1998}%
  \BibitemOpen
  \bibfield  {author} {\bibinfo {author} {\bibfnamefont {D.}~\bibnamefont {Gottesman}},\ }\href {https://arxiv.org/abs/quant-ph/9807006} {\bibinfo {title} {The heisenberg representation of quantum computers}} (\bibinfo {year} {1998}),\ \Eprint {https://arxiv.org/abs/quant-ph/9807006} {arXiv:quant-ph/9807006 [quant-ph]} \BibitemShut {NoStop}%
\bibitem [{\citenamefont {Gottesman}(1997)}]{gottesman1997}%
  \BibitemOpen
  \bibfield  {author} {\bibinfo {author} {\bibfnamefont {D.}~\bibnamefont {Gottesman}},\ }\href {https://arxiv.org/abs/quant-ph/9705052} {\bibinfo {title} {Stabilizer codes and quantum error correction}} (\bibinfo {year} {1997}),\ \Eprint {https://arxiv.org/abs/quant-ph/9705052} {arXiv:quant-ph/9705052 [quant-ph]} \BibitemShut {NoStop}%
\bibitem [{\citenamefont {Marconi}\ \emph {et~al.}(2024)\citenamefont {Marconi}, \citenamefont {Alfaro-Bittner}, \citenamefont {Sarrazin}, \citenamefont {Giudici},\ and\ \citenamefont {Tredicce}}]{marconi2024}%
  \BibitemOpen
  \bibfield  {author} {\bibinfo {author} {\bibfnamefont {M.}~\bibnamefont {Marconi}}, \bibinfo {author} {\bibfnamefont {K.}~\bibnamefont {Alfaro-Bittner}}, \bibinfo {author} {\bibfnamefont {L.}~\bibnamefont {Sarrazin}}, \bibinfo {author} {\bibfnamefont {M.}~\bibnamefont {Giudici}},\ and\ \bibinfo {author} {\bibfnamefont {J.}~\bibnamefont {Tredicce}},\ }\bibfield  {title} {\bibinfo {title} {Critical slowing down in a real physical system},\ }\href {https://doi.org/https://doi.org/10.1016/j.chaos.2024.115218} {\bibfield  {journal} {\bibinfo  {journal} {Chaos, Solitons \& Fractals}\ }\textbf {\bibinfo {volume} {186}},\ \bibinfo {pages} {115218} (\bibinfo {year} {2024})}\BibitemShut {NoStop}%
\bibitem [{\citenamefont {Scheffer}\ \emph {et~al.}(2009)\citenamefont {Scheffer}, \citenamefont {Bascompte}, \citenamefont {Brock}, \citenamefont {Brovkin}, \citenamefont {Carpenter}, \citenamefont {Dakos}, \citenamefont {Held}, \citenamefont {Van~Nes}, \citenamefont {Rietkerk},\ and\ \citenamefont {Sugihara}}]{scheffer2009}%
  \BibitemOpen
  \bibfield  {author} {\bibinfo {author} {\bibfnamefont {M.}~\bibnamefont {Scheffer}}, \bibinfo {author} {\bibfnamefont {J.}~\bibnamefont {Bascompte}}, \bibinfo {author} {\bibfnamefont {W.~A.}\ \bibnamefont {Brock}}, \bibinfo {author} {\bibfnamefont {V.}~\bibnamefont {Brovkin}}, \bibinfo {author} {\bibfnamefont {S.~R.}\ \bibnamefont {Carpenter}}, \bibinfo {author} {\bibfnamefont {V.}~\bibnamefont {Dakos}}, \bibinfo {author} {\bibfnamefont {H.}~\bibnamefont {Held}}, \bibinfo {author} {\bibfnamefont {E.~H.}\ \bibnamefont {Van~Nes}}, \bibinfo {author} {\bibfnamefont {M.}~\bibnamefont {Rietkerk}},\ and\ \bibinfo {author} {\bibfnamefont {G.}~\bibnamefont {Sugihara}},\ }\bibfield  {title} {\bibinfo {title} {Early-warning signals for critical transitions},\ }\href {https://doi.org/10.1038/nature08227} {\bibfield  {journal} {\bibinfo  {journal} {Nature}\ }\textbf {\bibinfo {volume} {461}},\ \bibinfo {pages} {53} (\bibinfo {year} {2009})}\BibitemShut {NoStop}%
\bibitem [{\citenamefont {Van~Hove}(1954)}]{vanhove1954}%
  \BibitemOpen
  \bibfield  {author} {\bibinfo {author} {\bibfnamefont {L.}~\bibnamefont {Van~Hove}},\ }\bibfield  {title} {\bibinfo {title} {Time-{{Dependent Correlations}} between {{Spins}} and {{Neutron Scattering}} in {{Ferromagnetic Crystals}}},\ }\href {https://doi.org/10.1103/PhysRev.95.1374} {\bibfield  {journal} {\bibinfo  {journal} {Phys. Rev.}\ }\textbf {\bibinfo {volume} {95}},\ \bibinfo {pages} {1374} (\bibinfo {year} {1954})}\BibitemShut {NoStop}%
\bibitem [{\citenamefont {Li}\ \emph {et~al.}(2022)\citenamefont {Li}, \citenamefont {Luo}, \citenamefont {Wang}, \citenamefont {Xie}, \citenamefont {Liu}, \citenamefont {Hu}, \citenamefont {Chen}, \citenamefont {Yao},\ and\ \citenamefont {Pan}}]{li2022}%
  \BibitemOpen
  \bibfield  {author} {\bibinfo {author} {\bibfnamefont {X.}~\bibnamefont {Li}}, \bibinfo {author} {\bibfnamefont {X.}~\bibnamefont {Luo}}, \bibinfo {author} {\bibfnamefont {S.}~\bibnamefont {Wang}}, \bibinfo {author} {\bibfnamefont {K.}~\bibnamefont {Xie}}, \bibinfo {author} {\bibfnamefont {X.-P.}\ \bibnamefont {Liu}}, \bibinfo {author} {\bibfnamefont {H.}~\bibnamefont {Hu}}, \bibinfo {author} {\bibfnamefont {Y.-A.}\ \bibnamefont {Chen}}, \bibinfo {author} {\bibfnamefont {X.-C.}\ \bibnamefont {Yao}},\ and\ \bibinfo {author} {\bibfnamefont {J.-W.}\ \bibnamefont {Pan}},\ }\bibfield  {title} {\bibinfo {title} {Second sound attenuation near quantum criticality},\ }\href {https://www.science.org/doi/10.1126/science.abi4480} {\bibfield  {journal} {\bibinfo  {journal} {Science}\ }\textbf {\bibinfo {volume} {375}},\ \bibinfo {pages} {528} (\bibinfo {year} {2022})}\BibitemShut {NoStop}%
\bibitem [{\citenamefont {Qian}\ and\ \citenamefont {Wang}(2024)}]{qian2024a}%
  \BibitemOpen
  \bibfield  {author} {\bibinfo {author} {\bibfnamefont {D.}~\bibnamefont {Qian}}\ and\ \bibinfo {author} {\bibfnamefont {J.}~\bibnamefont {Wang}},\ }\bibfield  {title} {\bibinfo {title} {Steering-induced phase transition in measurement-only quantum circuits},\ }\href {https://doi.org/10.1103/PhysRevB.109.024301} {\bibfield  {journal} {\bibinfo  {journal} {Phys. Rev. B}\ }\textbf {\bibinfo {volume} {109}},\ \bibinfo {pages} {024301} (\bibinfo {year} {2024})}\BibitemShut {NoStop}%
\bibitem [{\citenamefont {Paviglianiti}\ \emph {et~al.}(2025)\citenamefont {Paviglianiti}, \citenamefont {Fresco}, \citenamefont {Silva}, \citenamefont {Spagnolo}, \citenamefont {Valenti},\ and\ \citenamefont {Carollo}}]{Paviglianiti2024}%
  \BibitemOpen
  \bibfield  {author} {\bibinfo {author} {\bibfnamefont {A.}~\bibnamefont {Paviglianiti}}, \bibinfo {author} {\bibfnamefont {G.~D.}\ \bibnamefont {Fresco}}, \bibinfo {author} {\bibfnamefont {A.}~\bibnamefont {Silva}}, \bibinfo {author} {\bibfnamefont {B.}~\bibnamefont {Spagnolo}}, \bibinfo {author} {\bibfnamefont {D.}~\bibnamefont {Valenti}},\ and\ \bibinfo {author} {\bibfnamefont {A.}~\bibnamefont {Carollo}},\ }\bibfield  {title} {\bibinfo {title} {Breakdown of measurement-induced phase transitions under information loss},\ }\href {https://doi.org/10.22331/q-2025-06-24-1781} {\bibfield  {journal} {\bibinfo  {journal} {Quantum}\ }\textbf {\bibinfo {volume} {9}},\ \bibinfo {pages} {1781} (\bibinfo {year} {2025})}\BibitemShut {NoStop}%
\bibitem [{\citenamefont {Garratt}\ and\ \citenamefont {Altman}(2024)}]{garratt2024}%
  \BibitemOpen
  \bibfield  {author} {\bibinfo {author} {\bibfnamefont {S.~J.}\ \bibnamefont {Garratt}}\ and\ \bibinfo {author} {\bibfnamefont {E.}~\bibnamefont {Altman}},\ }\bibfield  {title} {\bibinfo {title} {Probing postmeasurement entanglement without postselection},\ }\href {https://journals.aps.org/prxquantum/abstract/10.1103/PRXQuantum.5.030311} {\bibfield  {journal} {\bibinfo  {journal} {PRX Quantum}\ }\textbf {\bibinfo {volume} {5}},\ \bibinfo {pages} {030311} (\bibinfo {year} {2024})}\BibitemShut {NoStop}%
\bibitem [{\citenamefont {McGinley}(2024)}]{mcginley2024}%
  \BibitemOpen
  \bibfield  {author} {\bibinfo {author} {\bibfnamefont {M.}~\bibnamefont {McGinley}},\ }\bibfield  {title} {\bibinfo {title} {Postselection-{{Free Learning}} of {{Measurement-Induced Quantum Dynamics}}},\ }\href {https://doi.org/10.1103/PRXQuantum.5.020347} {\bibfield  {journal} {\bibinfo  {journal} {PRX Quantum}\ }\textbf {\bibinfo {volume} {5}},\ \bibinfo {pages} {020347} (\bibinfo {year} {2024})}\BibitemShut {NoStop}%
\bibitem [{\citenamefont {Tikhanovskaya}\ \emph {et~al.}(2024)\citenamefont {Tikhanovskaya}, \citenamefont {Lavasani}, \citenamefont {Fisher},\ and\ \citenamefont {Vijay}}]{tikhanovskaya2024}%
  \BibitemOpen
  \bibfield  {author} {\bibinfo {author} {\bibfnamefont {M.}~\bibnamefont {Tikhanovskaya}}, \bibinfo {author} {\bibfnamefont {A.}~\bibnamefont {Lavasani}}, \bibinfo {author} {\bibfnamefont {M.~P.}\ \bibnamefont {Fisher}},\ and\ \bibinfo {author} {\bibfnamefont {S.}~\bibnamefont {Vijay}},\ }\bibfield  {title} {\bibinfo {title} {Universality of the cross-entropy in $\mathbb{Z}_2$ symmetric monitored quantum circuits},\ }\href {https://journals.aps.org/prb/abstract/10.1103/PhysRevB.109.224313} {\bibfield  {journal} {\bibinfo  {journal} {Phys. Rev. B}\ }\textbf {\bibinfo {volume} {109}},\ \bibinfo {pages} {224313} (\bibinfo {year} {2024})}\BibitemShut {NoStop}%
\bibitem [{\citenamefont {Kamakari}\ \emph {et~al.}(2025)\citenamefont {Kamakari}, \citenamefont {Sun}, \citenamefont {Li}, \citenamefont {Thio}, \citenamefont {Gujarati}, \citenamefont {Fisher}, \citenamefont {Motta},\ and\ \citenamefont {Minnich}}]{kamakari2024}%
  \BibitemOpen
  \bibfield  {author} {\bibinfo {author} {\bibfnamefont {H.}~\bibnamefont {Kamakari}}, \bibinfo {author} {\bibfnamefont {J.}~\bibnamefont {Sun}}, \bibinfo {author} {\bibfnamefont {Y.}~\bibnamefont {Li}}, \bibinfo {author} {\bibfnamefont {J.~J.}\ \bibnamefont {Thio}}, \bibinfo {author} {\bibfnamefont {T.~P.}\ \bibnamefont {Gujarati}}, \bibinfo {author} {\bibfnamefont {M.~P.~A.}\ \bibnamefont {Fisher}}, \bibinfo {author} {\bibfnamefont {M.}~\bibnamefont {Motta}},\ and\ \bibinfo {author} {\bibfnamefont {A.~J.}\ \bibnamefont {Minnich}},\ }\bibfield  {title} {\bibinfo {title} {Experimental demonstration of scalable cross-entropy benchmarking to detect measurement-induced phase transitions on a superconducting quantum processor},\ }\href {https://doi.org/10.1103/PhysRevLett.134.120401} {\bibfield  {journal} {\bibinfo  {journal} {Phys. Rev. Lett.}\ }\textbf {\bibinfo {volume} {134}},\ \bibinfo {pages} {120401} (\bibinfo {year} {2025})}\BibitemShut {NoStop}%
\bibitem [{\citenamefont {Holevo}(1998)}]{holevo1998}%
  \BibitemOpen
  \bibfield  {author} {\bibinfo {author} {\bibfnamefont {A.~S.}\ \bibnamefont {Holevo}},\ }\bibfield  {title} {\bibinfo {title} {The capacity of the quantum channel with general signal states},\ }\href {https://ieeexplore.ieee.org/document/651037} {\bibfield  {journal} {\bibinfo  {journal} {IEEE Trans. Inf. Theory}\ }\textbf {\bibinfo {volume} {44}},\ \bibinfo {pages} {269} (\bibinfo {year} {1998})}\BibitemShut {NoStop}%
\bibitem [{\citenamefont {Devetak}\ and\ \citenamefont {Shor}(2005)}]{devetak2005}%
  \BibitemOpen
  \bibfield  {author} {\bibinfo {author} {\bibfnamefont {I.}~\bibnamefont {Devetak}}\ and\ \bibinfo {author} {\bibfnamefont {P.~W.}\ \bibnamefont {Shor}},\ }\bibfield  {title} {\bibinfo {title} {The {{Capacity}} of a {{Quantum Channel}} for {{Simultaneous Transmission}} of {{Classical}} and {{Quantum Information}}},\ }\href {https://doi.org/10.1007/s00220-005-1317-6} {\bibfield  {journal} {\bibinfo  {journal} {Commun. Math. Phys.}\ }\textbf {\bibinfo {volume} {256}},\ \bibinfo {pages} {287} (\bibinfo {year} {2005})}\BibitemShut {NoStop}%
\bibitem [{\citenamefont {Hastings}(2009)}]{hastings2009}%
  \BibitemOpen
  \bibfield  {author} {\bibinfo {author} {\bibfnamefont {M.~B.}\ \bibnamefont {Hastings}},\ }\bibfield  {title} {\bibinfo {title} {Superadditivity of communication capacity using entangled inputs},\ }\href {https://doi.org/10.1038/nphys1224} {\bibfield  {journal} {\bibinfo  {journal} {Nature Phys.}\ }\textbf {\bibinfo {volume} {5}},\ \bibinfo {pages} {255} (\bibinfo {year} {2009})}\BibitemShut {NoStop}%
\bibitem [{\citenamefont {Devetak}(2005)}]{devetak2005a}%
  \BibitemOpen
  \bibfield  {author} {\bibinfo {author} {\bibfnamefont {I.}~\bibnamefont {Devetak}},\ }\bibfield  {title} {\bibinfo {title} {The {{Private Classical Capacity}} and {{Quantum Capacity}} of a {{Quantum Channel}}},\ }\href {https://doi.org/10.1109/TIT.2004.839515} {\bibfield  {journal} {\bibinfo  {journal} {IEEE Trans. Inf. Theory}\ }\textbf {\bibinfo {volume} {51}},\ \bibinfo {pages} {44} (\bibinfo {year} {2005})}\BibitemShut {NoStop}%
\bibitem [{\citenamefont {Lloyd}(1997)}]{lloyd1997}%
  \BibitemOpen
  \bibfield  {author} {\bibinfo {author} {\bibfnamefont {S.}~\bibnamefont {Lloyd}},\ }\bibfield  {title} {\bibinfo {title} {Capacity of the noisy quantum channel},\ }\href {https://doi.org/10.1103/PhysRevA.55.1613} {\bibfield  {journal} {\bibinfo  {journal} {Phys. Rev. A}\ }\textbf {\bibinfo {volume} {55}},\ \bibinfo {pages} {1613} (\bibinfo {year} {1997})}\BibitemShut {NoStop}%
\bibitem [{\citenamefont {DiVincenzo}\ \emph {et~al.}(1998)\citenamefont {DiVincenzo}, \citenamefont {Shor},\ and\ \citenamefont {Smolin}}]{divincenzo1998}%
  \BibitemOpen
  \bibfield  {author} {\bibinfo {author} {\bibfnamefont {D.~P.}\ \bibnamefont {DiVincenzo}}, \bibinfo {author} {\bibfnamefont {P.~W.}\ \bibnamefont {Shor}},\ and\ \bibinfo {author} {\bibfnamefont {J.~A.}\ \bibnamefont {Smolin}},\ }\bibfield  {title} {\bibinfo {title} {Quantum-channel capacity of very noisy channels},\ }\href {https://doi.org/10.1103/PhysRevA.57.830} {\bibfield  {journal} {\bibinfo  {journal} {Phys. Rev. A}\ }\textbf {\bibinfo {volume} {57}},\ \bibinfo {pages} {830} (\bibinfo {year} {1998})}\BibitemShut {NoStop}%
\bibitem [{\citenamefont {Barnum}\ \emph {et~al.}(2000)\citenamefont {Barnum}, \citenamefont {Knill},\ and\ \citenamefont {Nielsen}}]{barnum2000a}%
  \BibitemOpen
  \bibfield  {author} {\bibinfo {author} {\bibfnamefont {H.}~\bibnamefont {Barnum}}, \bibinfo {author} {\bibfnamefont {E.}~\bibnamefont {Knill}},\ and\ \bibinfo {author} {\bibfnamefont {M.}~\bibnamefont {Nielsen}},\ }\bibfield  {title} {\bibinfo {title} {On quantum fidelities and channel capacities},\ }\href {https://doi.org/10.1109/18.850671} {\bibfield  {journal} {\bibinfo  {journal} {IEEE Trans. Inf. Theory}\ }\textbf {\bibinfo {volume} {46}},\ \bibinfo {pages} {1317} (\bibinfo {year} {2000})}\BibitemShut {NoStop}%
\bibitem [{\citenamefont {Delfosse}\ and\ \citenamefont {Zémor}(2020)}]{delfosse2020}%
  \BibitemOpen
  \bibfield  {author} {\bibinfo {author} {\bibfnamefont {N.}~\bibnamefont {Delfosse}}\ and\ \bibinfo {author} {\bibfnamefont {G.}~\bibnamefont {Zémor}},\ }\bibfield  {title} {\bibinfo {title} {Linear-time maximum likelihood decoding of surface codes over the quantum erasure channel},\ }\href {https://doi.org/10.1103/PhysRevResearch.2.033042} {\bibfield  {journal} {\bibinfo  {journal} {Phys. Rev. Research}\ }\textbf {\bibinfo {volume} {2}},\ \bibinfo {pages} {033042} (\bibinfo {year} {2020})}\BibitemShut {NoStop}%
\bibitem [{\citenamefont {Solanki}\ and\ \citenamefont {Sarvepalli}(2023)}]{solanki2023}%
  \BibitemOpen
  \bibfield  {author} {\bibinfo {author} {\bibfnamefont {H.~M.}\ \bibnamefont {Solanki}}\ and\ \bibinfo {author} {\bibfnamefont {P.~K.}\ \bibnamefont {Sarvepalli}},\ }\bibfield  {title} {\bibinfo {title} {Decoding topological subsystem color codes over the erasure channel using gauge fixing},\ }\href {https://ieeexplore.ieee.org/document/10129188} {\bibfield  {journal} {\bibinfo  {journal} {IEEE Trans. Commun.}\ }\textbf {\bibinfo {volume} {71}},\ \bibinfo {pages} {4181} (\bibinfo {year} {2023})}\BibitemShut {NoStop}%
\bibitem [{\citenamefont {Grassl}\ \emph {et~al.}(1997)\citenamefont {Grassl}, \citenamefont {Beth},\ and\ \citenamefont {Pellizzari}}]{grassl1997}%
  \BibitemOpen
  \bibfield  {author} {\bibinfo {author} {\bibfnamefont {M.}~\bibnamefont {Grassl}}, \bibinfo {author} {\bibfnamefont {T.}~\bibnamefont {Beth}},\ and\ \bibinfo {author} {\bibfnamefont {T.}~\bibnamefont {Pellizzari}},\ }\bibfield  {title} {\bibinfo {title} {Codes for the {{Quantum Erasure Channel}}},\ }\href {https://doi.org/10.1103/PhysRevA.56.33} {\bibfield  {journal} {\bibinfo  {journal} {Phys. Rev. A}\ }\textbf {\bibinfo {volume} {56}},\ \bibinfo {pages} {33} (\bibinfo {year} {1997})}\BibitemShut {NoStop}%
\bibitem [{\citenamefont {Bennett}\ \emph {et~al.}(1997)\citenamefont {Bennett}, \citenamefont {DiVincenzo},\ and\ \citenamefont {Smolin}}]{bennett1997}%
  \BibitemOpen
  \bibfield  {author} {\bibinfo {author} {\bibfnamefont {C.~H.}\ \bibnamefont {Bennett}}, \bibinfo {author} {\bibfnamefont {D.~P.}\ \bibnamefont {DiVincenzo}},\ and\ \bibinfo {author} {\bibfnamefont {J.~A.}\ \bibnamefont {Smolin}},\ }\bibfield  {title} {\bibinfo {title} {Capacities of {{Quantum Erasure Channels}}},\ }\href {https://doi.org/10.1103/PhysRevLett.78.3217} {\bibfield  {journal} {\bibinfo  {journal} {Phys. Rev. Lett.}\ }\textbf {\bibinfo {volume} {78}},\ \bibinfo {pages} {3217} (\bibinfo {year} {1997})}\BibitemShut {NoStop}%
\bibitem [{\citenamefont {Chang}\ \emph {et~al.}(2024)\citenamefont {Chang}, \citenamefont {Singh}, \citenamefont {Claes}, \citenamefont {Sahay}, \citenamefont {Teoh},\ and\ \citenamefont {Puri}}]{chang2024}%
  \BibitemOpen
  \bibfield  {author} {\bibinfo {author} {\bibfnamefont {K.}~\bibnamefont {Chang}}, \bibinfo {author} {\bibfnamefont {S.}~\bibnamefont {Singh}}, \bibinfo {author} {\bibfnamefont {J.}~\bibnamefont {Claes}}, \bibinfo {author} {\bibfnamefont {K.}~\bibnamefont {Sahay}}, \bibinfo {author} {\bibfnamefont {J.}~\bibnamefont {Teoh}},\ and\ \bibinfo {author} {\bibfnamefont {S.}~\bibnamefont {Puri}},\ }\href {https://arxiv.org/abs/2408.00842} {\bibinfo {title} {Surface code with imperfect erasure checks}} (\bibinfo {year} {2024}),\ \Eprint {https://arxiv.org/abs/2408.00842} {arXiv:2408.00842 [quant-ph]} \BibitemShut {NoStop}%
\bibitem [{\citenamefont {Kang}\ \emph {et~al.}(2023)\citenamefont {Kang}, \citenamefont {Campbell},\ and\ \citenamefont {Brown}}]{kang2023}%
  \BibitemOpen
  \bibfield  {author} {\bibinfo {author} {\bibfnamefont {M.}~\bibnamefont {Kang}}, \bibinfo {author} {\bibfnamefont {W.~C.}\ \bibnamefont {Campbell}},\ and\ \bibinfo {author} {\bibfnamefont {K.~R.}\ \bibnamefont {Brown}},\ }\bibfield  {title} {\bibinfo {title} {Quantum {{Error Correction}} with {{Metastable States}} of {{Trapped Ions Using Erasure Conversion}}},\ }\href {https://doi.org/10.1103/PRXQuantum.4.020358} {\bibfield  {journal} {\bibinfo  {journal} {PRX Quantum}\ }\textbf {\bibinfo {volume} {4}},\ \bibinfo {pages} {020358} (\bibinfo {year} {2023})}\BibitemShut {NoStop}%
\bibitem [{\citenamefont {Wu}\ \emph {et~al.}(2022)\citenamefont {Wu}, \citenamefont {Kolkowitz}, \citenamefont {Puri},\ and\ \citenamefont {Thompson}}]{wu2022a}%
  \BibitemOpen
  \bibfield  {author} {\bibinfo {author} {\bibfnamefont {Y.}~\bibnamefont {Wu}}, \bibinfo {author} {\bibfnamefont {S.}~\bibnamefont {Kolkowitz}}, \bibinfo {author} {\bibfnamefont {S.}~\bibnamefont {Puri}},\ and\ \bibinfo {author} {\bibfnamefont {J.~D.}\ \bibnamefont {Thompson}},\ }\bibfield  {title} {\bibinfo {title} {Erasure conversion for fault-tolerant quantum computing in alkaline earth {{Rydberg}} atom arrays},\ }\href {https://doi.org/10.1038/s41467-022-32094-6} {\bibfield  {journal} {\bibinfo  {journal} {Nature Commun.}\ }\textbf {\bibinfo {volume} {13}},\ \bibinfo {pages} {4657} (\bibinfo {year} {2022})}\BibitemShut {NoStop}%
\bibitem [{\citenamefont {Kubica}\ \emph {et~al.}(2023)\citenamefont {Kubica}, \citenamefont {Haim}, \citenamefont {Vaknin}, \citenamefont {Levine}, \citenamefont {Brandão},\ and\ \citenamefont {Retzker}}]{kubica2023}%
  \BibitemOpen
  \bibfield  {author} {\bibinfo {author} {\bibfnamefont {A.}~\bibnamefont {Kubica}}, \bibinfo {author} {\bibfnamefont {A.}~\bibnamefont {Haim}}, \bibinfo {author} {\bibfnamefont {Y.}~\bibnamefont {Vaknin}}, \bibinfo {author} {\bibfnamefont {H.}~\bibnamefont {Levine}}, \bibinfo {author} {\bibfnamefont {F.}~\bibnamefont {Brandão}},\ and\ \bibinfo {author} {\bibfnamefont {A.}~\bibnamefont {Retzker}},\ }\bibfield  {title} {\bibinfo {title} {Erasure {{Qubits}}: {{Overcoming}} the {{T}} 1 {{Limit}} in {{Superconducting Circuits}}},\ }\href {https://doi.org/10.1103/PhysRevX.13.041022} {\bibfield  {journal} {\bibinfo  {journal} {Phys. Rev. X}\ }\textbf {\bibinfo {volume} {13}},\ \bibinfo {pages} {041022} (\bibinfo {year} {2023})}\BibitemShut {NoStop}%
\bibitem [{\citenamefont {Gu}\ \emph {et~al.}(2024)\citenamefont {Gu}, \citenamefont {Vaknin}, \citenamefont {Retzker},\ and\ \citenamefont {Kubica}}]{gu2024}%
  \BibitemOpen
  \bibfield  {author} {\bibinfo {author} {\bibfnamefont {S.}~\bibnamefont {Gu}}, \bibinfo {author} {\bibfnamefont {Y.}~\bibnamefont {Vaknin}}, \bibinfo {author} {\bibfnamefont {A.}~\bibnamefont {Retzker}},\ and\ \bibinfo {author} {\bibfnamefont {A.}~\bibnamefont {Kubica}},\ }\href {https://arxiv.org/abs/2408.00829} {\bibinfo {title} {Optimizing quantum error correction protocols with erasure qubits}} (\bibinfo {year} {2024}),\ \Eprint {https://arxiv.org/abs/2408.00829} {arXiv:2408.00829 [quant-ph]} \BibitemShut {NoStop}%
\bibitem [{\citenamefont {Collins}\ and\ \citenamefont {Sniady}(2006)}]{collins2006}%
  \BibitemOpen
  \bibfield  {author} {\bibinfo {author} {\bibfnamefont {B.}~\bibnamefont {Collins}}\ and\ \bibinfo {author} {\bibfnamefont {P.}~\bibnamefont {Sniady}},\ }\bibfield  {title} {\bibinfo {title} {Integration with respect to the {{Haar}} measure on unitary, orthogonal and symplectic group},\ }\href {https://doi.org/10.1007/s00220-006-1554-3} {\bibfield  {journal} {\bibinfo  {journal} {Commun. Math. Phys.}\ }\textbf {\bibinfo {volume} {264}},\ \bibinfo {pages} {773} (\bibinfo {year} {2006})}\BibitemShut {NoStop}%
\end{thebibliography}
\end{document}

\title{Supplemental Material for ``Coherent Information Phase Transition in a Noisy Quantum Circuit''}
\author{Dongheng Qian}
\affiliation{State Key Laboratory of Surface Physics and Department of Physics, Fudan University, Shanghai 200433, China}
\affiliation{Shanghai Research Center for Quantum Sciences, Shanghai 201315, China}
\author{Jing Wang}
\thanks{wjingphys@fudan.edu.cn}
\affiliation{State Key Laboratory of Surface Physics and Department of Physics, Fudan University, Shanghai 200433, China}
\affiliation{Shanghai Research Center for Quantum Sciences, Shanghai 201315, China}
\affiliation{Institute for Nanoelectronic Devices and Quantum Computing, Fudan University, Shanghai 200433, China}
\affiliation{Hefei National Laboratory, Hefei 230088, China}

%\date{\today}

\maketitle

\tableofcontents

\section{Analytic mapping}

\begin{figure}[t]
\begin{center}
\includegraphics[width=6.5in, clip=true]{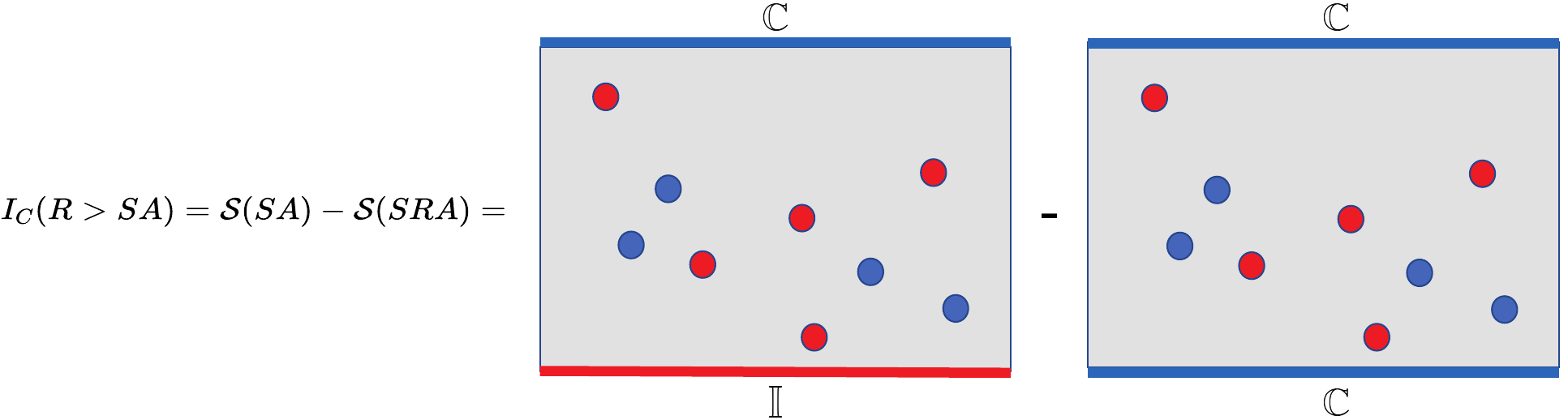}
\end{center}
\caption{Statistical mapping of the coherent information $I_C(R>SA)$. $I_C(R>SA)$ is represented by the free energy difference in the same random field model, where the bottom boundary condition changes from $\mathbb{I}$ to $\mathbb{C}$.}
\label{fig1}
\end{figure}

In this section, we explain in detail how the coherent information $I_C(R>SA)$ can be mapped to the energy difference of a statistical mechanics model under different boundary conditions. Moreover, the corresponding statistical mechanics model is ferromagnetic, and noise and QE operations are mapped to two competing external fields. Our derivation is largely based on Refs.~\cite{bao2020, jian2020a, zhou2019, dias2023, liu2023e, weinstein2022} and a direct generalization of Ref.~\cite{qian2024b}.

We label measurement trajectory by $m$ and denote $\rho_m = \left | \Psi_m \right \rangle \left \langle \Psi_m \right |$ as the purified state with support on the environment, system and ancilla qubits corresponding to the trajectory $m$. Notice that the dependence on circuit realization $C$ is omitted for brevity. This state is unnormalized, with the Born probability of occurrence given by $p_m = \text{Tr}(\rho_m)$. Given the initial state $\rho$, we have $\rho_m = C_m \rho C_m^{\dagger}$, where $C_m$ denotes all the random unitary gates, measurements, noises and QE operations. By definition, we have:
\begin{eqnarray}
 \overline{I_C(R>SA)} = \lim_{n \to 1}\overline{\mathcal{S}^{(n)}(SA)}-\overline{\mathcal{S}^{n}(SRA)} = \lim_{n \to 1}\mathop{\mathbb{E}}\limits_{\text{circuits}} \sum_{m}p_m\frac{1}{1-n}\text{log}\left[\frac{\text{Tr}(\rho_{SA,m}^{n})}{\text{Tr}(\rho_{SRA,m}^n)}\right],
\end{eqnarray}
where $ \rho_{SA, m} = \text{Tr}_{RE}(\rho_m)$ and $\rho_{SRA,m}=\text{Tr}_{E}(\rho_m)$. We denote $\mathbb{S}$ to be the cyclic permutation operation in permutation group $\mathbb{S}(n)$, and define its action on the $n$-fold replicated single qubit Hilbert space as:
\begin{equation}
\mathbb{S} = \sum_{i_1, i_2, ..i_n}\left | i_2,i_3, ..i_n,i_1\right \rangle \left \langle i_1,i_2, ..i_{n-1},i_n\right |.
\end{equation}
We can further define $\mathbb{S}^{X} = \otimes_{i \in X}\mathbb{S}_i$. Using the identities $ \text{Tr}(\rho_{X,m}^{n}) = \text{Tr}(\rho_m^{\otimes n}\mathbb{S}^{X})$ and $ \text{log}x=\lim_{k \to 0} \frac{x^{k}-1}{k}$, we obtain:
\begin{eqnarray}
\overline{I_C(R>SA)}=&&\lim_{n \to 1}\lim_{k \to 0}\mathop{\mathbb{E}}\limits_{\text{circuits}}\sum_{m}\frac{1}{(1-n)k}\text{Tr}(\rho_m)\left[\text{Tr}(\rho_m^{\otimes n}\mathbb{S}^{SR})^{k}-\text{Tr}(\rho_m^{\otimes n}\mathbb{S}^{SRA})^{k}
\right] \nonumber \\ 
=&& \lim_{n \to 1}\lim_{k \to 0}\mathop{\mathbb{E}}\limits_{\text{circuits}}\sum_{m}\frac{1}{(1-n)k}\left[\text{Tr}(\rho_m^{\otimes Q}\mathbb{C}^{SA}) - \text{Tr}(\rho_m^{\otimes Q}\mathbb{C}^{SRA})\right].
\end{eqnarray}
Here, $\mathbb{C}^{X} = \otimes_{i \in X}(\mathbb{S}^{\otimes k}\otimes I)$ and $Q=nk+1$. We can define $\Lambda_{A}^{Q}=\text{Tr}_{A\cup E}[\rho_m^{Q}\mathbb{C}^{A}]$, and the above equation eventually becomes:
\begin{eqnarray}
\overline{I_C(R>SA)}=&&\lim_{n \to 1}\lim_{k \to 0}  \mathop{\mathbb{E}}\limits_{\text{circuits}}\sum_{m}\frac{1}{(1-n)k}[\text{Tr}_{SR}(\Lambda_{A}^{Q}\mathbb{C}^{S}) - \text{Tr}_{SR}(\Lambda_{A}^{Q}\mathbb{C}^{SR})] = \lim_{n \to 1}\lim_{k \to 0} \frac{1}{(1-n)k}(\mathcal{Z}_{S}^{(Q)}-\mathcal{Z}_{SR}^{(Q)}) \\
&&\mathcal{Z}_{SA}^{(Q)} =\mathop{\mathbb{E}}\limits_{\text{circuits}}\sum_{m}\text{Tr}_{SR}\left(\mathbb{C}^{S}\Lambda_{A}^{(Q)}\right), \mathcal{Z}_{SRA}^{(Q)} = \mathop{\mathbb{E}}\limits_{\text{circuits}}\sum_{m}\text{Tr}_{SR}\left(\mathbb{C}^{SR}\Lambda_{A}^{(Q)}\right).
\end{eqnarray}
Notice that $\lim_{k\to 0}\mathcal{Z}_{SA}^{(Q)} = \mathcal{Z}_{SA}^{(1)}=1$ and $\lim_{k\to 0}\mathcal{Z}_{SRA}^{(Q)} = \mathcal{Z}_{SRA}^{(1)}=1$. Therefore, we have:
\begin{equation}
\lim_{k \to 0}(\mathcal{Z}_{SA}^{(Q)}-\mathcal{Z}_{SRA}^{(Q)}) = \lim_{k\to0}\text{log}\left(\frac{\mathcal{Z}_{SA}^{(Q)}}{\mathcal{Z}_{SRA}^{(Q)}}\right)
\end{equation},
which recovers the results in the main text. We can further employ a convenient vector notation where the replicated density matrix $\Lambda_{A}^{Q}$ is regarded as a vector state $\left | \Lambda_{A}^{Q} \right \rangle$ in the $ \mathcal{H}^{\otimes Q}\otimes \mathcal{H}^{*\otimes Q}$ Hilbert space. Notice that this state is supported on both system qubits $S$ and reference qubits $R$. We define states corresponding to a particular group member $g$ in the permutation group $S(Q)$ as:
\begin{equation}
\left | g \right \rangle = \sum_{i_1,i_2,...i_{Q}}^{d}\left | i_{g(1)},i_{g(2)}, ...i_{g(Q)};i_{1},i_{2},...i_{Q} \right \rangle,
\label{eq:S8}
\end{equation}
where $d$ is the local Hilbert dimension. Thus, we have 
$\mathcal{Z}_{SA}^{Q} =  \left \langle \mathcal{Z}_{\text{bulk}} |\ \mathbb{C}^{S}\otimes \mathbb{I}^{R} \right \rangle$
and
$\mathcal{Z}_{SRA}^{Q} =  \left \langle \mathcal{Z}_{\text{bulk}} |\ \mathbb{C}^{S\cup R} \right \rangle$
\normalsize
with $\left | \mathcal{Z}_{\text{bulk}} \right \rangle = \mathop{\mathbb{E}}\limits_{\text{circuits}}\sum_m\left |\Lambda_{A}^{(Q)} \right \rangle$. The  goal is then to represent $\left | \mathcal{Z}_{\text{bulk}} \right \rangle$ as a statistical mechanics model for spins taking value in permutation group $\mathbb{S}_{Q}$, with $\mathcal{Z}_{SA}^{Q}$ and $\mathcal{Z}_{SRA}^{Q}$ being the partition function with the boundary conditions $\left | \mathbb{C}^{S}\otimes \mathbb{I}^{R}\right \rangle$ and $\left | \mathbb{I}^{S\cup R} \right \rangle$, respectively. It is important to note that $S$ denotes the upper boundary, while $R$ corresponds to the lower boundary, as illustrated in Fig.~\ref{fig1}.

With the notation in Eq.~(\ref{eq:S8}), we can represent various operations in the circuit in an intuitive and simple manner. Two identities are particularly useful: $\left \langle \sigma | i,..i;i,..i\right \rangle = 1 $ and $\left \langle \sigma | \tau \right \rangle = d^{|\sigma \tau^{-1}|}$, where $|g|$ denotes the number of cycles in permutation $g$. It’s evident to see that this inner product reaches its maximum when $\sigma = \tau$, which suggests that it acts like $d^{Q}\delta_{\sigma, \tau}$ in the limit $d \to \infty$. The random unitary gates contribute to the vertical bond weight, given by:
\begin{equation}
\mathbb{E}_{\text{Haar}}[U_{ij}^{\otimes Q}\otimes U_{ij}^{*\otimes Q}] = \sum_{\{\sigma, \tau\}\in \mathbb{S}(Q)}\text{Wg}^{Q}(d^{2};\sigma^{-1}\tau)\left | \sigma\sigma\right \rangle_{ij}\left \langle \tau \tau \right |_{ij}.
\end{equation}
Here, $\text{Wg}^{Q}(d^{2};g)$ is the Weingarten function for the unitary group $U(d^{2})$~\cite{collins2006}. In the $d \to \infty$ limit, it behaves like:
\begin{equation}
\lim_{d \to \infty} \text{Wg}^{(Q)}(d^2;\sigma^{-1}\tau)\sim\frac{\text{Moeb}(\sigma^{-1}\tau)}{d^{4Q-2|\sigma^{-1}\tau|}},
\label{s9}
\end{equation}
where $\text{Moeb}(\sigma^{-1}\tau)$ is the Moebius number of permutation $\sigma^{-1}\tau$. Thus, it acts as the ferromagnetic coupling along the vertical bonds which aligns $\sigma$ and $\tau$.
Other operations correspond to the weight $W$ on non-vertical bonds. Consequently, the statistical mechanics model is an anisotropic $Q!$ state model on a honeycomb lattice, with the partition function given by:
\begin{equation}
\label{eq:s10}
\mathcal{Z}_{\text{bulk}} = \sum_{\{\sigma,\tau\} \in S(Q)}\prod_{\left \langle\sigma \tau\right \rangle \in {V^{c}}}W(\sigma, \tau)\prod_{\left \langle\sigma \tau\right \rangle \in {V}}\text{Wg}^{(Q)}(d^{2};\sigma^{-1}\tau),
\end{equation}
where $V$ denotes all the vertical bonds and $V^{c}$ denotes the non-vertical bonds.

For a projective measurement with outcome $i$, the weight is $\left \langle \sigma | i,..i;i,..i\right \rangle \left \langle i,..i;i,..i | \tau \right \rangle = 1$. Since we further average over all the measurement outcomes, the measurement contributes a weight $d$ on the bond.  For the QE operation considered in this work, we take $U_{SA}=\text{SWAP}$ and $\rho_A =  \left | 0 \right \rangle \left \langle 0 \right |$. This contributes a weight of $\left \langle \sigma | \mathbb{C} \right \rangle $, which acts as an external field pointing in the direction of $\mathbb{C}$. Conversely, noise acts as an external field pointing in the direction of $\mathbb{I}$. For example, in the case of depolarizing noise, where $U_{SE}=\text{SWAP}, \rho_{E}=\frac{1}{\sqrt{2}}(\left | 00 \right \rangle + \left | 11 \right \rangle)(\left \langle 00 \right | + \left \langle 11 \right |) $, the total bond weight for non-vertical bonds is given by:
\begin{eqnarray}
    W^{\text{depo}}(\sigma, \tau) = (1-p)(1-q_t)\left \langle \sigma | \tau \right \rangle + dp(1-qq_t)  + qq_t [(1-p)\left \langle \sigma | \mathbb{I} \right \rangle+dp] \left \langle \mathbb{I} | \tau \right \rangle  + (1-p)(1-q)q_t\left \langle \sigma | \mathbb{C}\right \rangle.
\end{eqnarray}
The results for resetting noise and dephasing noise can be derived in a similar way:
\begin{eqnarray}
W^{\text{reset}}(\sigma, \tau) &=& (1-p)(1-q_t)\left \langle \sigma | \tau \right \rangle + (1-p)qq_t\left \langle \sigma | \mathbb{I}\right \rangle + (1-p)(1-q)q_t\left \langle \sigma | \mathbb{C}\right \rangle + pd, \\ \nonumber
W^{\text{deph}}(\sigma, \tau) &\simeq  & (1-p)(1-q_t)\left \langle \sigma | \tau \right \rangle + (1-p)qq_t\left \langle \sigma | \mathbb{I}\right \rangle \delta_{\sigma, \tau} + (1-p)(1-q)q_t\left \langle \sigma | \mathbb{C}\right \rangle + pdqq_t\delta_{\sigma, \tau} + pd(q-qq_t),
\end{eqnarray}
where $W^{\text{deph}}(\sigma, \tau)$ is approximated in the limit $d \to \infty$. It is straightforward to observe that the effect of depolarizing noise is stronger than that of resetting noise, as it aligns both $\left | \sigma \right \rangle$ and $\left | \tau \right \rangle$ with $\left | \mathbb{I} \right \rangle$, while resetting noise only aligns $\left | \sigma \right \rangle$  with $\left | \mathbb{I} \right \rangle$. Conversely, dephasing noise is weaker than resetting noise for finite $d$, as corrections of order $O(1/d)$ would lead to alignment lead to alignment in directions other than $\mathbb{I}$~\cite{qian2024b}.

We now further examine the symmetry of the random field model with non-vertical bond weight. The symmetry group of the statistical mechanics model in the absence of noise and QE operation is $ (\mathbb{S}_{Q}\times \mathbb{S}_{Q}) \rtimes \mathbb{Z}_2$. The $\mathbb{S}_{Q}\times \mathbb{S}_{Q}$ symmetry is from the invariance of the weights under transformation $ \sigma \to g_i\sigma g_j$, where $g_i, g_j \in \mathbb{S}_{Q}$. The $\mathbb{Z}_2$ symmetry $\sigma \to \sigma^{-1}$ is due to the hermiticity of the density matrix.  When a net external field is present, the original $\mathbb{S}_{Q}\times \mathbb{S}_{Q}$ symmetry is explicitly broken and the symmetry group becomes $ \mathcal{C}_{\mathbb{S}_{Q}}(\mathbb{C})\rtimes \mathbb{Z}_{2}$. The $\mathbb{Z}_2$ symmetry remains to be $\sigma \to \sigma^{-1}$. The group $\mathcal{C}_{\mathbb{S}_{Q}}(\mathbb{C})$ is the subgroup of $\mathbb{S}_{Q}$ that stabilizes $\mathbb{C}$, meaning that any group element $g$ satisfies $[g, \mathbb{C}] = 0$. The corresponding symmetry operation is then $\sigma \to g\sigma g^{-1}$.

Finally, it is important to highlight an additional criterion required to formally validate $\mathcal{Z}_{\text{bulk}}$ as a proper partition function, that is, the positivity of all associated weights. However, it is well-known that the Weingarten function can assume negative values. This complication has already been observed in the original random quantum circuits, which do not incorporate measurements, noise or QE operations~\cite{zhou2019}. Previous studies have identified two specific scenarios under which positivity of weights is guaranteed: when $Q=2$ and in the limit $d \rightarrow \infty$~\cite{bao2020}. In the following, we show that the inclusion of noise and QE operations preserves weight positivity within both regimes.
First, considering the limit $d \rightarrow \infty$, the leading term of the Weingarten function according to Eq.~(\ref{s9}) is given by $\text{Wg}^{(Q)}(d^2; \sigma^{-1}\tau) \approx d^{-2Q}\delta_{\sigma,\tau} + O(d^{-2Q-(2Q-|\sigma^{-1}\tau|)})$, which is explicitly positive. Given that all weights associated with non-vertical bonds are non-negative, it follows that the total weight of any configuration remains non-negative in this limit.
Second, we analyze the case $Q=2$. Here, the permutation group consists solely of two elements: $\mathbb{I}$ and $\mathbb{C}$. Previous works have shown that weights can be rendered non-negative upon integrating out the $\tau$ spins~\cite{bao2020}. Specifically, considering a downward triangle configuration, with spins $\sigma_{1}, \sigma_{2}, \sigma_{3}$ located at the top-left, top-right, and bottom vertices respectively, and the spin $\tau$ positioned centrally, we aim to compute the following weight:
\begin{equation}
W'(\sigma_{1}, \sigma_{2}, \sigma_{3}) = \sum_{\tau \in \{\mathbb{I}, \mathbb{C}\}}W(\sigma_{1},\tau)W(\sigma_{2},\tau)\text{Wg}^{2}(d^2;\sigma_{3}^{-1}\tau)
\end{equation}
Owing to the symmetry properties $\sigma_{1} \leftrightarrow \sigma_{2}$ and $\mathbb{I} \leftrightarrow \mathbb{C}, q \leftrightarrow 1-q$, we only need to verify positivity for three distinct configurations: $W'(\mathbb{I}, \mathbb{I}, \mathbb{I})$, $W'(\mathbb{I}, \mathbb{I}, \mathbb{C})$, and $W'(\mathbb{I}, \mathbb{C}, \mathbb{I})$, given the parameter ranges $0 \leq p,q,q_{t} \leq 1$ and $d \geq 2$. Taking resetting noise as an illustrative example, we have explicitly:
\begin{equation}
W'(\mathbb{I}, \mathbb{I}, \mathbb{I})=\frac{1}{d^4-1}A(A+1+(d-1)(1-p)(1-q)q_{t}),
\end{equation}
where $A \equiv (d-1)\left((d+1)(1-p)(1-(1-q)q_{t})+p\right)$, which is manifestly non-negative.
Moreover, we obtain:
\begin{equation}
W'(\mathbb{I}, \mathbb{I},\mathbb{C})=\frac{1}{d^4-1}(C+D)((d^2-1)(1-p)(1-q)q_{t}+p(d-1)),
\end{equation}
with $C=d+d(d-1)(1-p)(1-q)q_{t}$ and $D=d(1-p)-(d-1)(1-p)(1-q)q_{t}+p$, both of which are non-negative.
Finally, we have:
\begin{eqnarray}
W'(\mathbb{I},\mathbb{C}, \mathbb{I})&=&\frac{1}{(d+1)(d^2+1)}(1+(d-1)(1-p)(1-q)q_{t}) \\ \nonumber
& \times & (d+p+(1-p)qq_{t}+d^2(1-p)(1-(1-q)q_{t})),
\end{eqnarray}
which is clearly non-negative as well. Therefore, integrating out the $\tau$ spins yields a partition function defined on a triangular lattice with non-negative three-body weights. When $q_{t}=0$ and $p=1$, these derived expressions precisely match those presented in Ref.~\cite{bao2020}. 

\section{Resource efficient protocol}

\begin{figure}[b]
\begin{center}
\includegraphics[width=6.5in, clip=true]{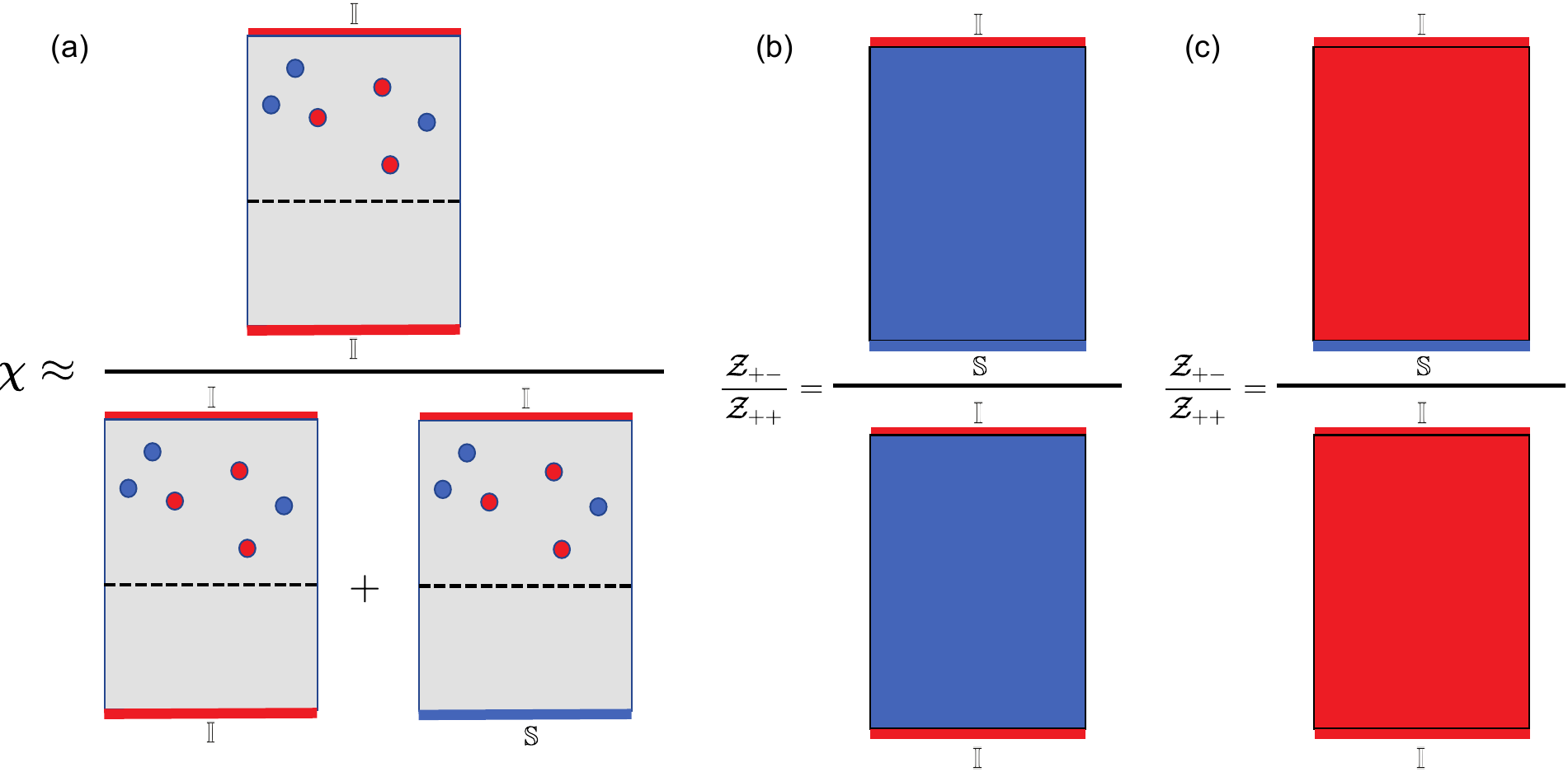}
\end{center}
\caption{Statistical mapping of $\chi$. (a) $\chi$ can be approximated as the quotient of partition functions under different boundary conditions on the bottom boundary. Regions below the dashed line representing the encoding stage. (b) The ratio $\mathcal{Z}_{+-}/\mathcal{Z}_{++}$ in the recoverable phase, where QE operations dominate. (c) Same as (b), but for the irrecoverable phase. }
\label{fig2}
\end{figure}

In this section, we provide an analytical explanation for why $\chi$ can be used to directly detect the coherent information phase transition. We will demonstrate that $\chi$ corresponds to the quotient of partition functions of a random field Ising model under different boundary conditions. To begin, we approximate $\chi$ using the annealed average:
\begin{equation}
\chi = \mathbb{E}_{\mathcal{C}}\frac{\sum_{m}\text{Tr}\left(\rho_{A}^{m} \sigma_{A}^{m}\right)}{\sum_{m}\text{Tr}\left[(\sigma_{A}^{m})^{2}\right]} \approx \frac{\mathbb{E}_{\mathcal{C}}\sum_{m}\text{Tr}\left(\rho_{A}^{m} \sigma_{A}^{m}\right)}{\mathbb{E}_{\mathcal{C}}\sum_{m}\text{Tr}\left[(\sigma_{A}^{m})^{2}\right]} = \frac{\mathbb{E}_{\mathcal{C}}\sum_{m}\text{Tr}\left[C_m^{\otimes2}(\rho\otimes\sigma)C_m^{\dagger \otimes 2}\mathbb{S}^{A}\right]}{\mathbb{E}_{\mathcal{C}}\sum_{m}\text{Tr}\left[C_m^{\otimes2}(\sigma\otimes\sigma)C_m^{\dagger \otimes 2}\mathbb{S}^{A}\right]}.
\end{equation}
Here, $\mathbb{S}$ represents the swap operator acting on two replicas. It is worth noting that this expression reduces to the cross-entropy benchmark introduced in Ref.~\cite{li2023} when QE operations are absent. This approximation is valid when fluctuations between different realizations and trajectories are small. We can further rewrite $\chi$ as:
\begin{equation}
\chi \approx  \frac{\left \langle \mathbb{I}_{S} | \mathcal{Z}_{\rho \sigma} \right \rangle}{\left \langle \mathbb{I}_{S} | \mathcal{Z}_{\sigma \sigma} \right \rangle}, \mathcal{Z}_{\rho \sigma} = \text{Tr}_{A\cup E}\left [C_m^{\otimes 2}(\rho \otimes \sigma)C_m^{\dagger \otimes 2} \mathbb{S}^{A} \right ], \mathcal{Z}_{\sigma \sigma} = \text{Tr}_{A\cup E}\left [C_m^{\otimes 2}(\sigma \otimes \sigma)C_m^{\dagger \otimes 2} \mathbb{S}^{A} \right].
\end{equation}
Here, $\mathcal{Z}_{\rho \sigma}$ and $\mathcal{Z}_{\sigma \sigma}$ are partition functions, including contributions from the bottom boundary, while $\mathbb{I}_S$ denotes the boundary condition on the top boundary. We can further decompose the bulk and bottom boundary contributions as $\mathcal{Z}_{\rho \sigma} = \mathcal{Z}_{\text{bulk}} Z_{\rho \sigma}$ and $\mathcal{Z}_{\sigma \sigma} = \mathcal{Z}_{\text{bulk}} Z_{\sigma \sigma}$. Notably, $\mathcal{Z}_{\text{bulk}}$ is precisely the partition function of a random field Ising model as in Eq.~(\ref{eq:s10}). Thus, a first-order phase transition would occur when changing the direction of the net external field. Due to the encoding stage, the replica spins $s_x$ must be identical on the bottom boundary~\cite{li2023}. Assuming $\rho = \prod_{x} \rho_x$ and $\sigma = \prod_{x} \sigma_x$ with $\text{Tr}(\rho_x) = \text{Tr}(\sigma_x) = 1$, and further assuming that $\sigma$ is a pure state such that $\text{Tr}[(\sigma_x)^{2}] = 1$, we have:
\begin{eqnarray}
Z_{\rho \sigma} &=& \prod_{x} \delta_{s_x=\mathbb{I}}\text{Tr}(\rho_x)\text{Tr}(\sigma_x)+\delta_{s_x=\mathbb{S}}\text{Tr}(\rho_x \sigma_x) = \prod_{x}\delta_{s_x=\mathbb{I}}+\delta_{s_x=\mathbb{S}}\text{Tr}(\rho_x \sigma_x), \\
Z_{\sigma \sigma} &=& \prod_{x} \delta_{s_x=\mathbb{I}}(\text{Tr}(\sigma_x))^{2}+\delta_{s_x=\mathbb{S}}\text{Tr}[(\sigma_x)^{2}] = \prod_{x}\delta_{s_x=\mathbb{I}}+\delta_{s_x=\mathbb{S}}.
\end{eqnarray}
Since $\text{Tr}(\rho_x \sigma_x) < 1$ for distinct $\rho$ and $\sigma$, $Z_{\rho \sigma}$ imposes a boundary condition in which the spins align with $\mathbb{I}$, whereas $Z_{\sigma \sigma}$ corresponds to an open boundary condition. Therefore, we have:
\begin{equation}
\chi \approx \frac{\mathcal{Z}_{++}}{\mathcal{Z}_{++} + \mathcal{Z}_{+-}} = \frac{1}{1 + \mathcal{Z}_{+-} / \mathcal{Z}_{++}},
\end{equation}
where $+$ and $-$ denote shorthand for boundary conditions $\mathbb{I}$ and $\mathbb{C}$, respectively. This is also illustrated in Fig.~\ref{fig2}(a). 

In the recoverable phase, the bulk spins align with $\mathbb{S}$, as shown in Fig.~\ref{fig2}(b). In this regime, $\mathcal{Z}_{+-}/\mathcal{Z}_{++} \sim \text{exp}(O(L))$, leading to $\chi=0$ in the thermodynamic limit. On the other hand, in the irrecoverable phase depicted in Fig.~\ref{fig2}(c), $\mathcal{Z}_{+-}/\mathcal{Z}_{++} \sim \text{exp}(-O(L))$, resulting in $\chi = 1$. Thus, the coherent information phase transition is reflected by a sharp transition in $\chi$ from $0$ to $1$.

It is important to emphasize that potential noise during the encoding stage does not compromise the efficacy of this protocol, as such noise can be interpreted as simply a different choice of $\rho$. In fact, one can select $\rho$ to be the maximally mixed state, rendering the unitary operations and noise inconsequential to its evolution. However, while $\chi$ can be estimated without the complications of post-selection, one area that may benefit from improvement is the measurement process. Following classical simulation, the Pauli operators to be measured often exhibit extensive support on the ancilla qubits, requiring a complex measurement circuit. In practical experiments, the measurement outcomes are highly sensitive to noise within the circuit, which can introduce errors~\cite{gottesman2009}. Thus, it would be advantageous to explore alternative probes that could eliminate the need for measuring long Pauli operators, thereby simplifying the experimental setup.

\section{Numerical details}
\subsection{Efficient simulation}
In this section, we introduce the method we used in numerical calculation. We first review the stabilizer formalism used to achieve efficient numerical simulation~\cite{aaronson2004, gottesman1998, gottesman1997}. We denote the Pauli group on $n$ qubits, $G_n$, as the group consisting of all Pauli matrices ${I, X, Y, Z}$ along with multiplicative factors ${\pm1, \pm i}$. A stabilizer group $S$ is a subgroup of $G_n$ generated by elements $g_1,...,g_l$. These generators are independent commuting generators. We can associate a stabilizer state with every stabilizer group as: 
\begin{equation}
\rho = \sum_{g \in G_n}g = \frac{2^l}{2^{n}}\prod_{i=1}^{l}\frac{1+g_i}{2}.
\end{equation}
When $n=l$, the stabilizer state becomes a pure state. These states can be represented efficiently by associating two $l \times n$ binary matrices, assuming phase is not considered. Specifically, each stabilizer can be represented as $g = \prod_{i=1}^{n}X_i^{\alpha_i}Z_i^{\beta_i}$ up to a phase factor, where $\alpha_i, \beta_i$ are binary numbers taking value in ${0, 1}$. Thus, instead of storing $2^n$ complex numbers, any stabilizer state can be efficiently represented by only $O(n)$ numbers. 

We demonstrate that all operations in the circuit map states to other states with a stabilizer representation, enabling efficient simulation. The two-qubit unitary gates are randomly sampled from the Clifford group, which is generated by $\{\text{CNOT}, \text{SWAP}, \text{H}, \text{P}\}$, representing the CNOT, SWAP, Hadamard and phase gate, respectively. According to the Gottesman–Knill theorem~\cite{nielsen2010}, a quantum circuit consisting solely of Clifford gates and measurements represented by Pauli operators can be efficiently simulated.  Specifically,  Clifford gates have the property of being the normalizer of $G_n$, meaning that $g' = UgU^{\dagger}$ still lies within the stabilizer group. Thus, we can track the evolution of the stabilizers rather than the state itself. For a measurement represented by a Pauli operator $g$, there are three possibilities:
\begin{itemize}
\item $g$ is in the stabilizer. The measurement result is determined by the phase, and the state is unchanged. 
\item $g$ is not in the stabilizer but commutes with all stabilizers. The measurement outcome is random, and $g$ is added to the stabilizer generators.
\item $g$ is not in the stabilizer and anti-commutes with a single generator $g'$. The measurement outcome is random, and the original $g'$ is replaced by $g$.
\end{itemize}
In the last case, one can always ensure that only one generator anti-commutes with $g$ by performing Gaussian elimination.
As long as $\rho_n, \rho_e$ are stabilizer states and $U_{SE}, U_{SA}$ are in the Clifford group, the noise and QE operations are also efficiently simulated. Specifically for noise, the partial trace operation on the environment qubits is performed by eliminating all generators with non-trivial support on the environment qubits after Gaussian elimination. 

To calculate the entanglement, it has been shown that the entanglement entropy of a subsystem $M$ is:
\begin{equation}
\mathcal{S}_M = |M| - \left|G_M\right|,
\label{eq:one}
\end{equation}
where $|M|$ is the number of qubits in $M$ and $|G_M|$ is the number of generators for the stabilizer group who only have non-trivial support in $M$~\cite{li2019}.This can also be practically calculated by tracing out $M^c$, with the number of remaining non-identity generators corresponding to $|G_M|$.

Although the above method is already efficient, in practice, we may have to store $\sim O(L^4)$ ancilla qubits to incorporate the QE operation. This still poses a significant challenge to memory and gate operation efficiency. However, we can exploit the fact that every ancilla qubit is acted upon non-trivially only once, and we are ultimately only interested in the conditional entanglement entropy. This numerical method, first proposed in Ref.~\cite{kelly2024a}, is succinctly reviewed here. Each time an ancilla qubit is coupled to the system, we can perform Gaussian elimination and then discard stabilizers that have non-trivial support only on the ancilla qubit and eliminate the remaining stabilizer’s support on the ancilla qubit. This approach prevents the number of generators from increasing linearly with time, ensuring the memory complexity to be $O(L^2)$. It’s worth noticing that the remaining generators may not commute with each other since we doesn’t keep track of the possibly non-trivial part on the ancilla qubits.

We denote the total number of generators discarded by $x$. Notice that these generators won’t be updated during the subsequent evolution, so they remain being non-trivial only on the ancilla qubits. The formula for the conditional entanglement entropy $\mathcal{S}(M|A) = \mathcal{S}(M,A)-\mathcal{S}(A)$ is extremely simple in this construction. For $\mathcal{S}(M,A)$, we can traceout $M^{c}$ and denote the number of stabilizers that are not identity as $y$. Using Eq.~(\ref{eq:one}), we have $\mathcal{S}(M, A) = |M| + |A| - x - y$. On the other hand, $\mathcal{S}(A)$ is simply $|A| - x$ since every stabilizer with non-trivial support only on the ancilla qubits are already discarded. It follows that: 
\begin{equation}
\mathcal{S}(M|A) = |M| - y.
\end{equation}
Remarkably, the dependence on the number of ancilla qubits and the number of discarded generators cancels out, allowing the conditional entanglement entropy to be calculated using only the generators of the final system state. We can then express the coherent information through the conditional entanglement entropy as:
\begin{equation}
I_C(R>SA) = \mathcal{S}(S|A) - \mathcal{S}(SR|A),
\end{equation}
where we regard the total system to be $SR$ and $M = S, M^c = R$. 

In the numerical simulation of the resource-efficient protocol, it is crucial to emphasize that we must keep track of the operators acting on the ancilla qubits, as access to the final state $\sigma_A$ of the ancilla qubits is required. This imposes a limitation on the scale of our numerical simulation.

\subsection{Data collapse}
\begin{figure}[htbp]
\begin{center}
\includegraphics[width=6.5in, clip=true]{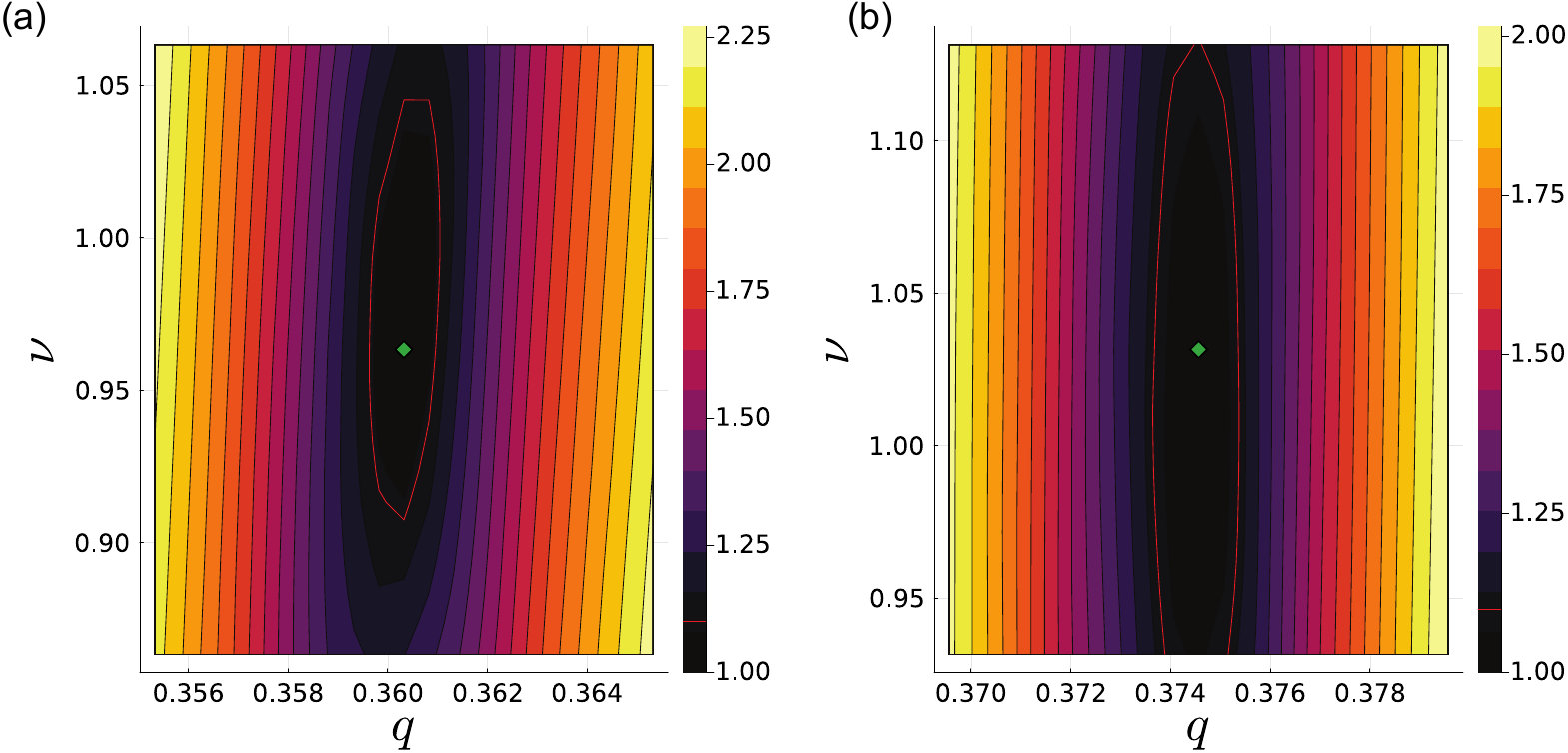}
\end{center}
\caption{Data collapse for $I_C$. The residues are rescaled by dividing the $\epsilon^{\text{min}}$. Red circle is where the residue equals $1.1\epsilon^{\text{min}}$ and the green diamond is where the minimal point is. (a) Depolarizing noise with $p=0.0$. (b) Depolarizing noise with $p=0.1$.}
\label{fig3}
\end{figure}

For both $I_C$ and $\chi$, we collapse the $I_3$ data according to the scaling form $y = f((p-p_c)L^{1/\nu})$. The data collapse procedure follows the method described in~\cite{qian2024, qian2024b} and proceeds as follows. For a given combination of $p_c$ and $\nu$, we rescale a particular data point $(p, L, I_C(\chi))$ as: 
\begin{equation}
    x = (p-p_c)L^{1/\nu},\ \ y=I_C(\chi).
\end{equation}
After rescaling all the data points, we fit the rescaled data with a $12$-th order polynomial and calculate the residue for the best fit. The residue $\epsilon(p_c, \nu)$ is then defined as the target function. By applying the Nelder-Mead algorithm, we find the minimal point $(p_c^{\text{min}}, \nu^{\text{min}})$ and the minimal residue $\epsilon^{\text{min}}$. To estimate the uncertainty in $p_c$ and $\nu$, we further plot the residue in the parameter space around the critical point. We set the threshold to be $1.1\epsilon^{\text{min}}$ to determine the uncertainty. Fig.~\ref{fig3} show examples for estimating the error for the depolarizing noise case with $p=0.0$ and $p=0.1$, respectively.

\subsection{More numerical results}

\begin{figure}[b]
\begin{center}
\includegraphics[width=6.5in, clip=true]{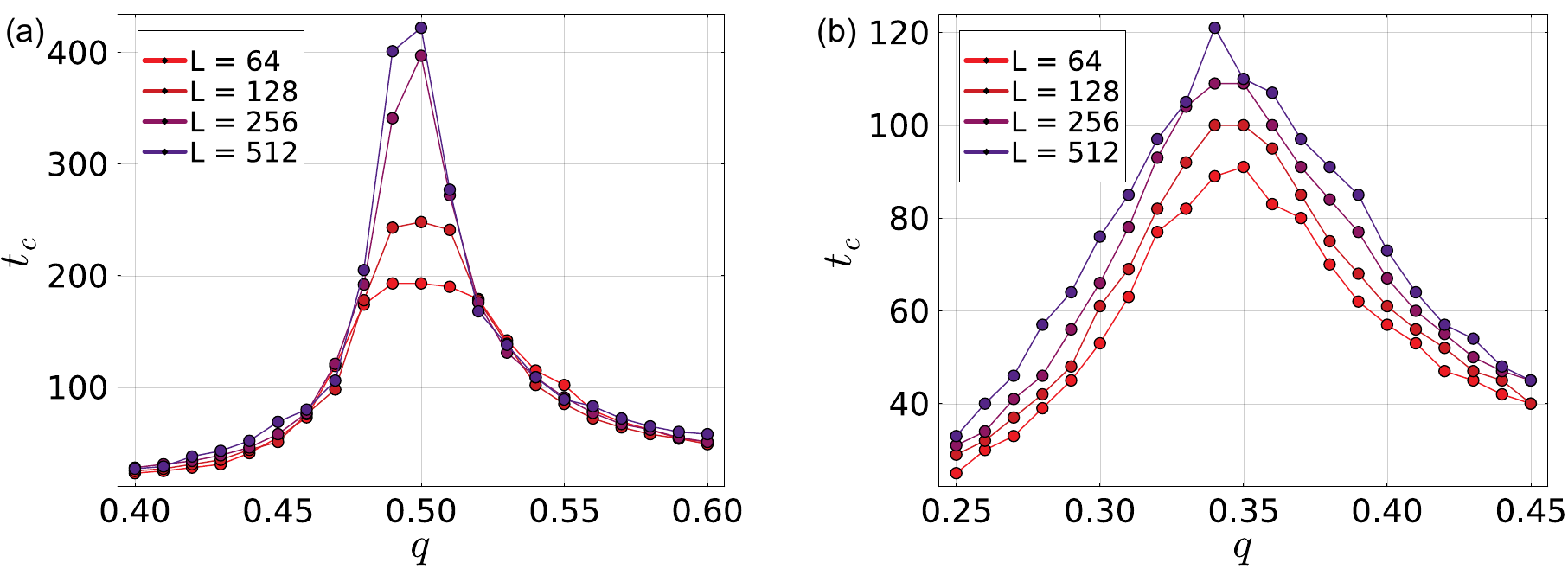}
\end{center}
\caption{Critical slowing down effect for other kinds of noises. (a) Dephasing noise. (b) Depolarizing noise.}
\label{fig4}
\end{figure}

In this section, we present numerical results on the coherent information phase transition, the critical slowing down effect, and the resource-efficient protocol across various noises and parameters. Additionally, it is noteworthy that the coherent information in the absence of noise was previously analyzed in Ref.~\cite{weinstein2023}. In that work, it was demonstrated that the coherent information remains positive, a result that is consistent with our findings for the regime deep within the recoverable phase.

We begin by demonstrating that the critical slowing down effect is also observed for other types of noise, as illustrated in Fig.~\ref{fig4}. Notably, the convergence time is significantly longer for dephasing noise compared to depolarizing and reset noise. The underlying cause of this difference still needs to be explored.

\begin{figure}[t]
\begin{center}
\includegraphics[width=6.5in, clip=true]{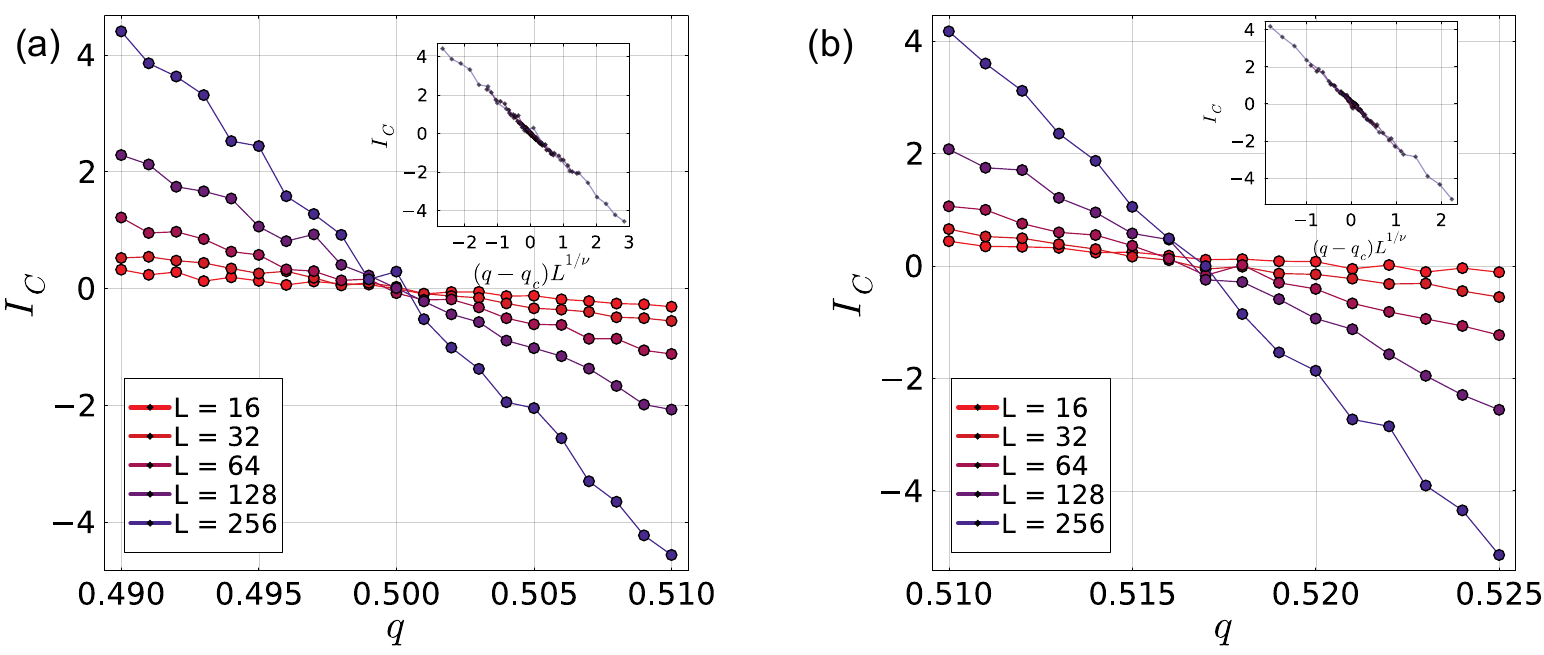}
\end{center}
\caption{Coherent information phase transition with $p=0.1$. Results for depolarizing noise has been shown in the main text. Insets are the data collapse results, by which critical points and exponents are determined.(a) Resetting noise. $q_c=0.499(1)$ and $\nu = 1.0(1)$. (b) Dephasing noise.  $q_c = 0.517(1)$ and $\nu=1.0(1)$.}
\label{fig5}
\end{figure}

\begin{figure}[t]
\begin{center}
\includegraphics[width=7.0in, clip=true]{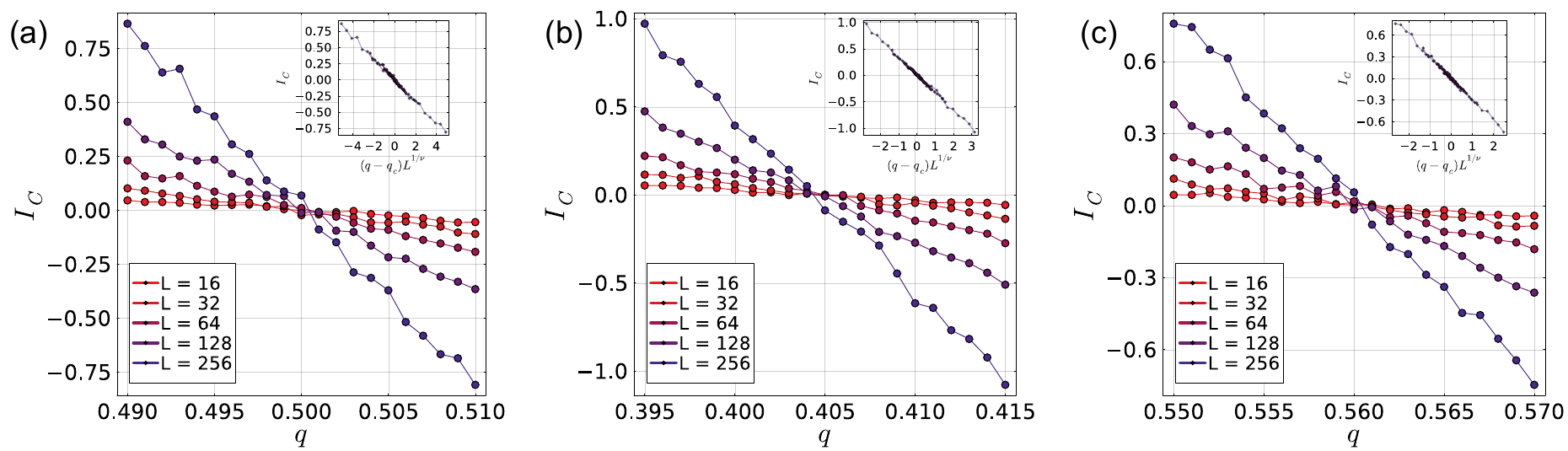}
\end{center}
\caption{Coherent information phase transition with $p=0.4$. Insets are the data collapse results, by which critical points and exponents are determined. (a) Resetting noise. $q_c=0.500(1)$ and $\nu=0.9(1)$. (b) Depolarizing noise. $q_c=0.403(1)$ and $\nu=1.1(1)$.(c) Dephasing noise. $q_c=0.559(2)$ and $\nu=1.0(2)$.}
\label{fig6}
\end{figure}

\begin{figure}[t]
\begin{center}
\includegraphics[width=7.0in, clip=true]{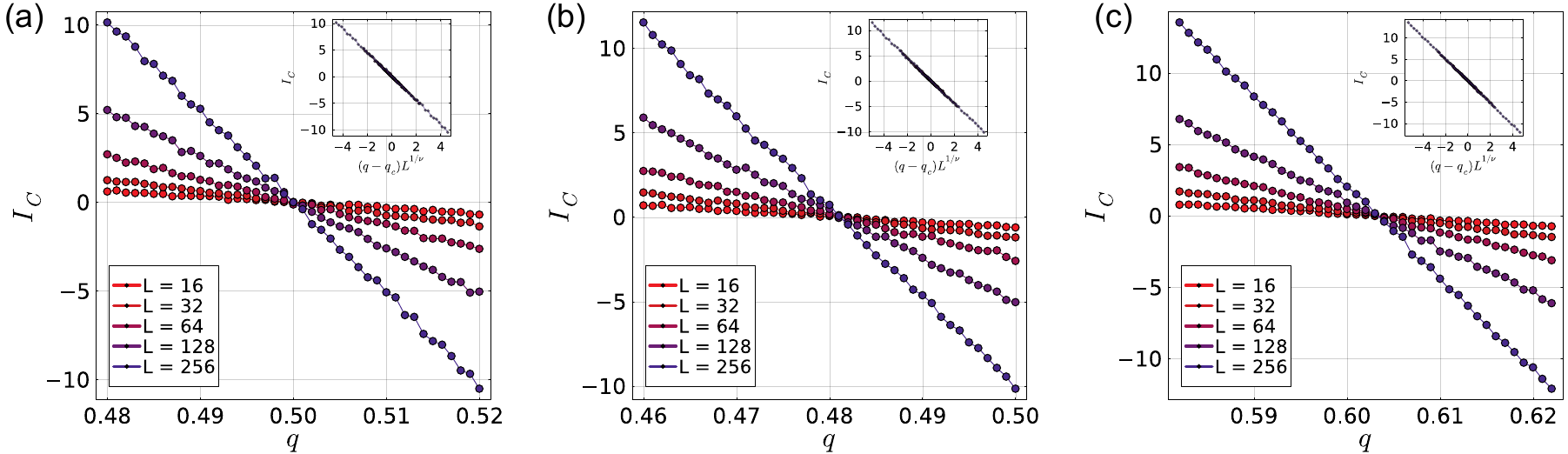}
\end{center}
\caption{Coherent information phase transition with $q_t=0.7$. Insets are the data collapse results, by which critical points and exponents are determined. (a) Resetting noise. $q_c=0.500(1)$ and $\nu=1.02(5)$. (b) Depolarizing noise. $q_c=0.481(1)$ and $\nu=1.01(5)$.(c) Dephasing noise. $q_c=0.603(1)$ and $\nu=1.01(2)$.}
\label{fig7}
\end{figure}

The coherent information phase transition occurs for any measurement rate $p < 1$ (when $p = 1$, every qubit is projectively measured, and $I_C = 0$ regardless of $q$). Results for $p = 0.1$ and $p = 0.4$ are shown in Fig.~\ref{fig5} and Fig.~\ref{fig6}, respectively. Furthermore, the phase transition is observed for any choice of $q_t$, as the primary effect of $q_t$ is to determine the total strength of the external field, whereas the crucial factor is the direction of the field, which is characterized by $q$. In the main text, we set $q_t = 0.1$, with the result for $q_t = 0.7$ displayed in Fig.~\ref{fig7}. It should be noted that for both resetting noise and depolarizing noise, the phase transition point $q_c$ approaches $0.5$ as $q_t \to 1$. This scenario corresponds to the case where half of the qubits are preserved while the other half are immediately discarded, a result that is ensured by the decoupling inequality~\cite{nielsen2010}.

\begin{figure}[t]
\begin{center}
\includegraphics[width=5.5in, clip=true]{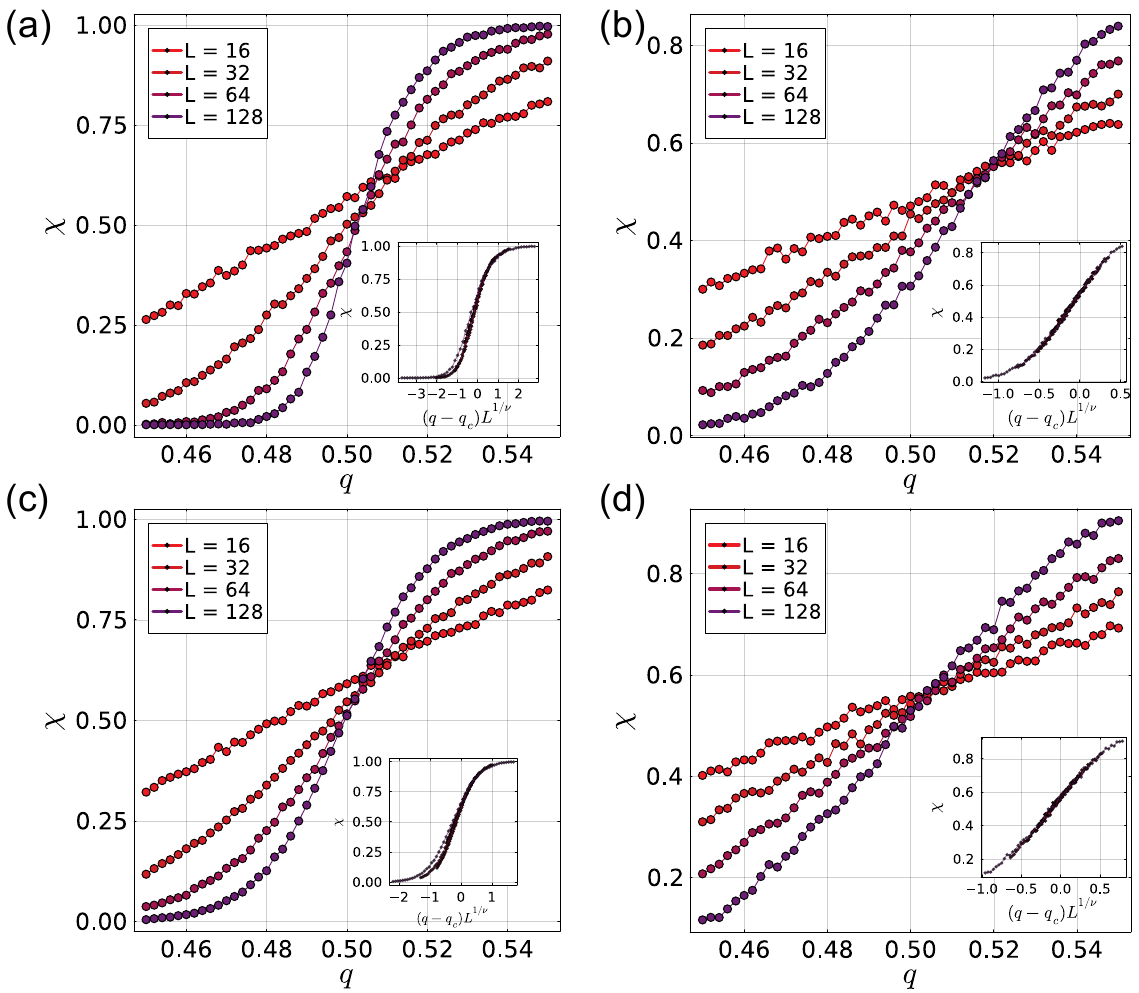}
\end{center}
\caption{Phase transition revealed by the efficient protocol. We choose $ \rho = (\left | + \right \rangle \left \langle + \right \rangle)^{\otimes L/2}\otimes(\left | 0 \right \rangle \left \langle 0 \right \rangle)^{\otimes L/2}$ and $ \sigma = (\left | 0 \right \rangle \left \langle 0 \right \rangle)^{\otimes L}$. Every data point is averaged over $3\times10^3$ realizations. The inset shows the data collapse result. (a) Dephasing noise with $p=0$. $q_c=0.507(2)$ and $\nu=1.2(1)$. (b) Dephasing noise with $p=0.1$. $q_c=0.519(3)$ and $\nu=1.7(1)$.(c) Resetting noise with $p=0$. $q_c=0.504(2)$ and $\nu=1.7(1)$. (d) Resetting noise with $p=0.1$. $q_c=0.506(3)$ and $\nu=1.3(1)$. }
\label{fig8}
\end{figure}

The resource-efficient protocol for detecting the coherent information phase transition is applicable for any measurement rate $p$ and any type of noise. The results for $p = 0.0$ and $p = 0.1$ under resetting and dephasing noise are shown in Fig.~\ref{fig8}. The critical points identified through data collapse are in agreement with the exact phase transition points.

\begin{figure}[t]
\begin{center}
\includegraphics[width=5.5in, clip=true]{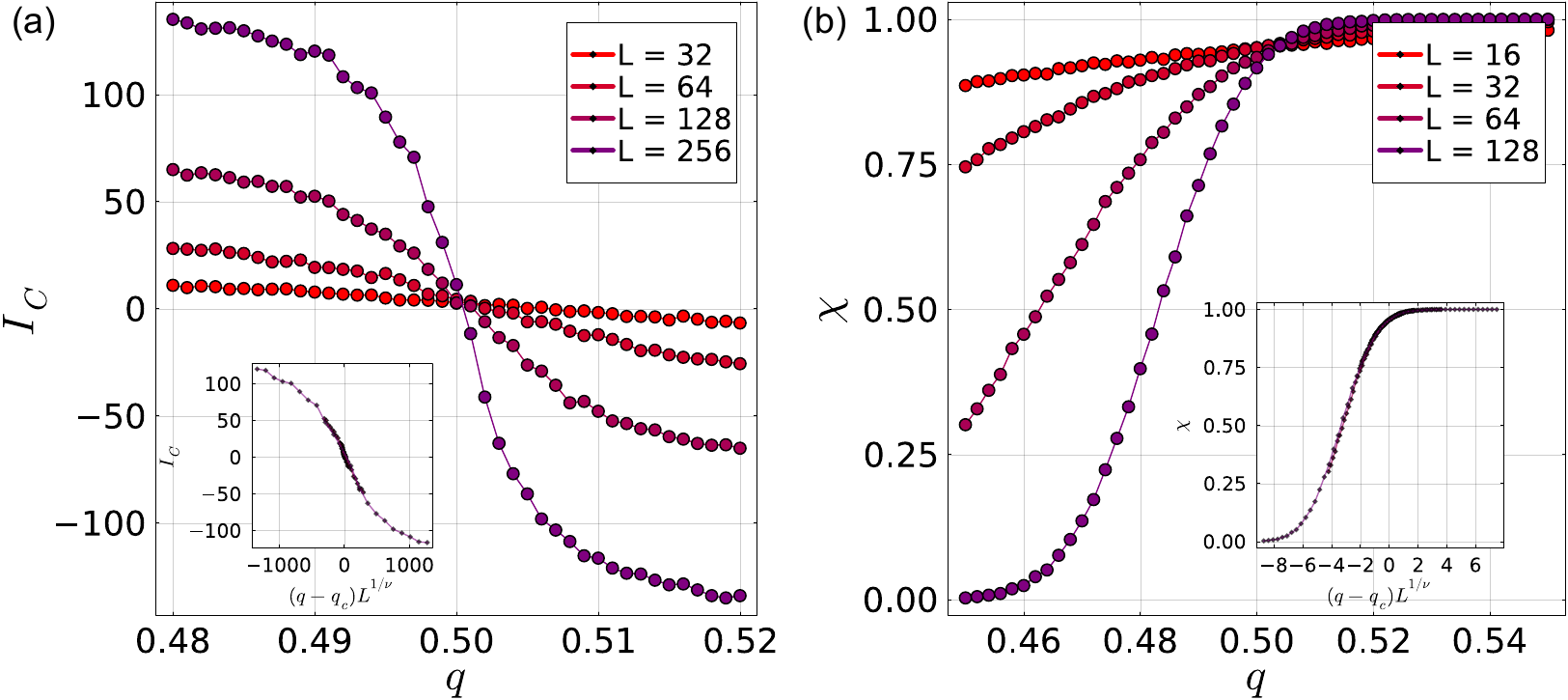}
\end{center}
\caption{Coherent information phase transition with $q_t = 0.03$ and the results revealed by the efficient protocol, considering resetting noise. Insets show the data collapse results, which are used to determine the critical points and exponents. (a) Coherent information phase transition, with $q_c = 0.500(1)$ and $\nu = 0.47(5)$. (b) Efficient protocol, with $q_c = 0.504(2)$ and $\nu = 1.0(2)$.}
\label{fig9}
\end{figure}

Another interesting case occurs when $q_t$ and $p$ are very small, where a volume law phase, corresponding to a ferromagnetic phase in the statistical mechanics model, emerges when the noise and QE operations are balanced~\cite{qian2024b}. In this regime, tuning $q$ directly corresponds to a first-order phase transition in the classical statistical mechanics model. Not only would the sign of $I_C$ change abruptly, but its value would also undergo a discontinuous jump at the coherent information phase transition point. We consider $q_t = 0.03$ and $p=0$ with resetting noise, and the results are shown in Fig.~\ref{fig9}(a), where it is evident that as the system size increases, $I_C$ experiences a sudden jump rather than a continuous drop for larger $q_t$. In this scenario, the efficient protocol still holds, as demonstrated in Fig.~\ref{fig9}(b).

\newpage

% \bibliography{supprefs}